\documentclass[notitlepage, preprint]{revtex4-1}
\usepackage{graphicx}
\usepackage{siunitx}
\usepackage{textcomp} %Upright Mu
\usepackage{amsmath}
\usepackage{bibentry}
\setcitestyle{super}

\usepackage{etoolbox}

\makeatletter
\patchcmd{\frontmatter@abstract@produce}
  {\vskip200\p@\@plus1fil
   \penalty-200\relax
   \vskip-200\p@\@plus-1fil}
  {}
  {}
\makeatother

\begin{document}

\title{Scanning Transmission Electron Microscopy \\ of Oxide Interfaces and Heterostructures}

\author{Steven R. Spurgeon}
\affiliation{Energy and Environment Directorate \\ Pacific Northwest National Laboratory, Richland, WA 99352, USA}

\date{\today}

\maketitle

\section{Introduction}

Thin film oxides are a source of endless fascination for the materials scientist. These materials are highly flexible, can be integrated into almost limitless combinations, and exhibit many useful functionalities for device applications. While precision synthesis techniques, such as molecular beam epitaxy (MBE) and pulsed laser deposition (PLD), provide a high degree of control over these systems, there remains a disconnect between ideal and realized materials. Because thin films adopt structures and chemistries distinct from their bulk counterparts, it is often difficult to predict what properties will emerge. The complex energy landscape of the synthesis process is also strongly influenced by non-equilibrium growth conditions imposed by the substrate, as well as the kinetics of thin film crystallization and fluctuations in process variables, all of which can lead to significant deviations from targeted outcomes.

High-resolution structural and chemical characterization techniques of the kind described in this volume, are needed to verify growth models, bound theoretical calculations, and guide materials design. While many characterization options exist, most are spatially-averaged or indirect, providing only partial insight into the complex behavior of these systems. Over the past several decades, scanning transmission electron microscopy (STEM) has become a cornerstone of oxide heterostructure characterization owing to its ability to simultaneously resolve structure, chemistry, and defects at the highest spatial resolution. STEM methods are an essential complement to averaged scattering techniques, offering a direct picture of resulting materials that can inform and refine the growth process to achieve targeted properties. There is arguably no other technique that can provide such a broad array of information at the atomic-scale, all within a single experimental session.

STEM analysis relies on the strong interaction of electrons with matter: a process that has been increasingly better understood, controlled, and applied since the development of the first electron microscope by Ruska and Knoll in 1931.\cite{Lambert1996} Over the past century, electron microscopy has moved from a niche tool to a pillar of materials characterization, a transformation enabled by the rise of modern computing, precision electronics, vacuum technologies, and advanced detectors. Today's instruments are a marvel of engineering, incorporating a particle accelerator that boosts electrons to nearly the speed of light, ultra-stable stages that drift a few nanometers over the course of a day, and cameras that can acquire thousands of frames per second with single electron sensitivity. Alongside the platform itself, the microscopy community has grown to encompass all fields of science, including physics, chemistry, biology, and medicine. The last decade alone has seen three Nobel prizes in chemistry and physics awarded for electron microscopy. The efforts of scientists worldwide have helped established highly effective workflows to prepare, image, and analyze a broad range of materials systems. STEM techniques are particularly well-suited to the examination of engineered complex oxides, which offer controlled structures and compositions, tolerance for high electron beam currents, and a rich theoretical framework for interpretation. The materials community's focus on directed materials design has strongly benefited from the virtuous cycle of synthesis, characterization, and modeling enabled by electron microscopy.

\section{Components of the Modern Electron Microscope}

\begin{figure}
\includegraphics[width=\textwidth]{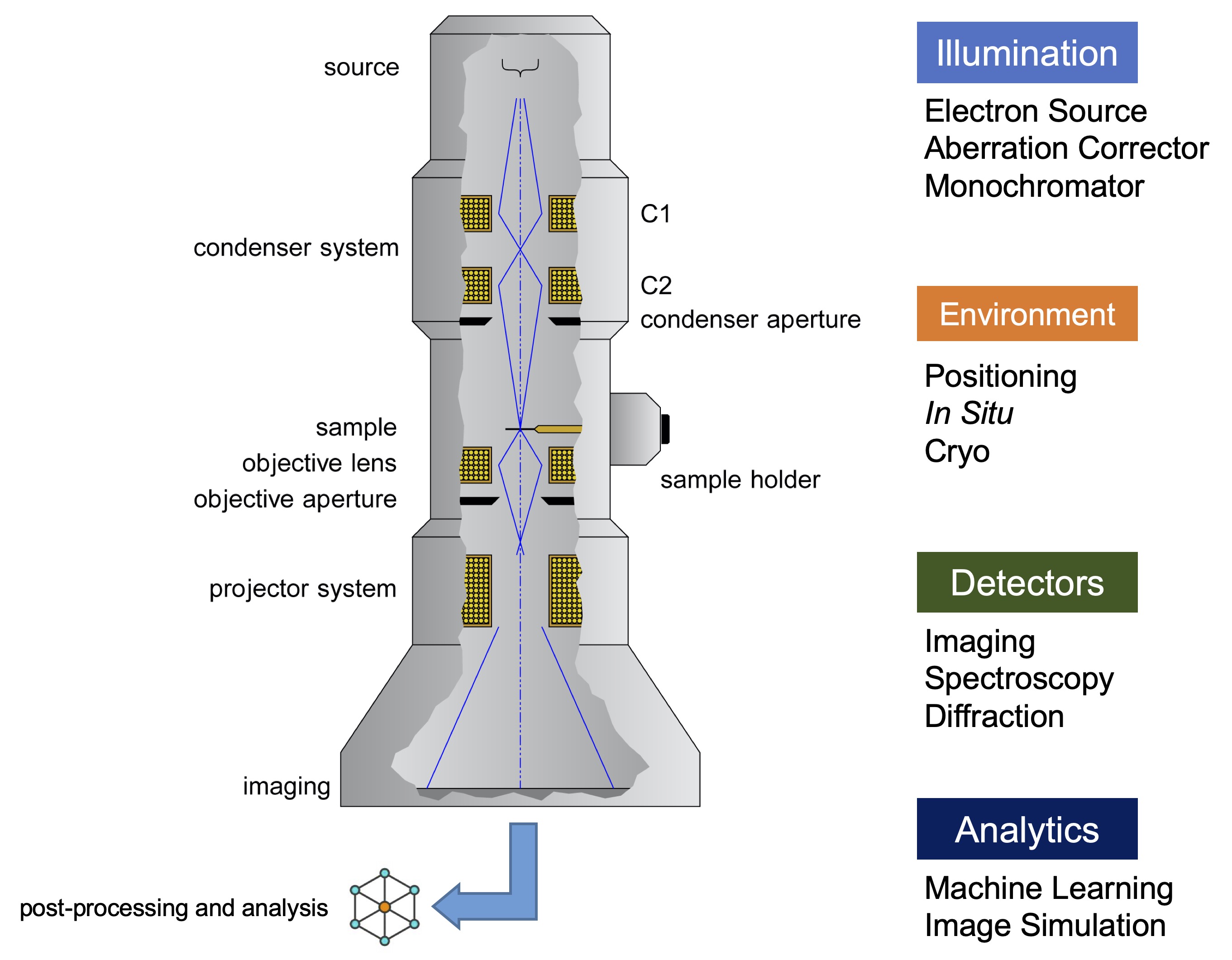}
\caption{Illustration of the scanning transmission electron microscope, marked with four major components of imaging and analysis. Adapted with permission of Dr. Eric Jensen.}
\label{scope_overview}
\end{figure}

A full appreciation of electron microscopy depends on understanding the evolution and present state of the many technologies inside today's instruments. We first summarize the operation, history, and ongoing developments in each major microscope component, then discuss their collective impact on selected problems related to thin film oxide interfaces and heterostructures. As shown in Figure \ref{scope_overview}, the microscope can be broadly divided into four components: the illumination system, sample environment, detectors, and data analytics. Our understanding of many fundamental processes, ranging from interface engineering to phase separation and point defect formation, has benefitted from improvements in the resolution and sensitivity of these components. Just as important, data is now collected and analyzed in new ways, decreasing the time it takes to get meaningful information from the microscope.

\subsection{Illumination System}\label{sec:illumination}

In the most common analytical configuration of the instrument, the upper portion of the column consists of an electron gun encompassing a tungsten ``cold'' field emission gun (FEG), to which a $\sim1$ kV potential is applied to induce electron tunneling. This source offers a high $10^{13}$ Am$^{-2}~$sr brightness and 0.3 eV energy spread compared to the $10^{11}$ Am$^{-2}~$sr brightness and 1.5 eV energy spread of a cheaper LaB$_6$ thermionic source at 100 kV.\cite{Williams2009} However, this source quality comes at the expense of a more stringent vacuum system, since the gun operates at a pressure of $10^{-11}$ Torr, as well as the need to periodically ``flash'' contamination from the tip---a process that is simple and quick on today's instruments. Electrons emitted from the tip are accelerated through a 60--300 kV potential (with more options available) and then passed through a series of lenses and apertures, the exact of configuration of which depends on whether the instrument is operating in parallel-beam TEM or convergent-beam STEM mode. The imaging process can be described in relatively simple terms as the interaction of an electron plane wave with a thin ($<100$ nm) foil sample. According to the weak phase object approximation, the effect of a phase object (sample) is to introduce a phase shift of the beam; that is, we can mathematically represent the object by a transmission function.\cite{Pennycook2011a} Manipulation of the mathematics describing this interaction (described in the volume by Pennycook) shows that the bright-field signal is coherent and can vary in contrast depending on the phase of the selected transfer function, whereas the dark-field signal is incoherent and does not show such variations. This property leads to the more direct interpretation of incoherent imaging such as high-angle annular dark-field (HAADF or ``$Z$-contrast'') and explains its widespread adoption by the community. While the exact choice of imaging mode depends on the features being examined and the desired contrast mechanism, STEM generally offers the highest spatial resolution for analytical imaging and will be the focus of this chapter. A broad range of techniques, including HAADF imaging, annular bright-field (ABF) imaging, energy-dispersive X-ray spectroscopy (EDS), and electron energy loss spectroscopy (EELS), are only possible or most effective in the STEM mode. These techniques have delivered unprecedented insights into the atomic structure, chemistry, and defects in complex oxides.

\begin{figure}
\includegraphics[width=0.65\textwidth]{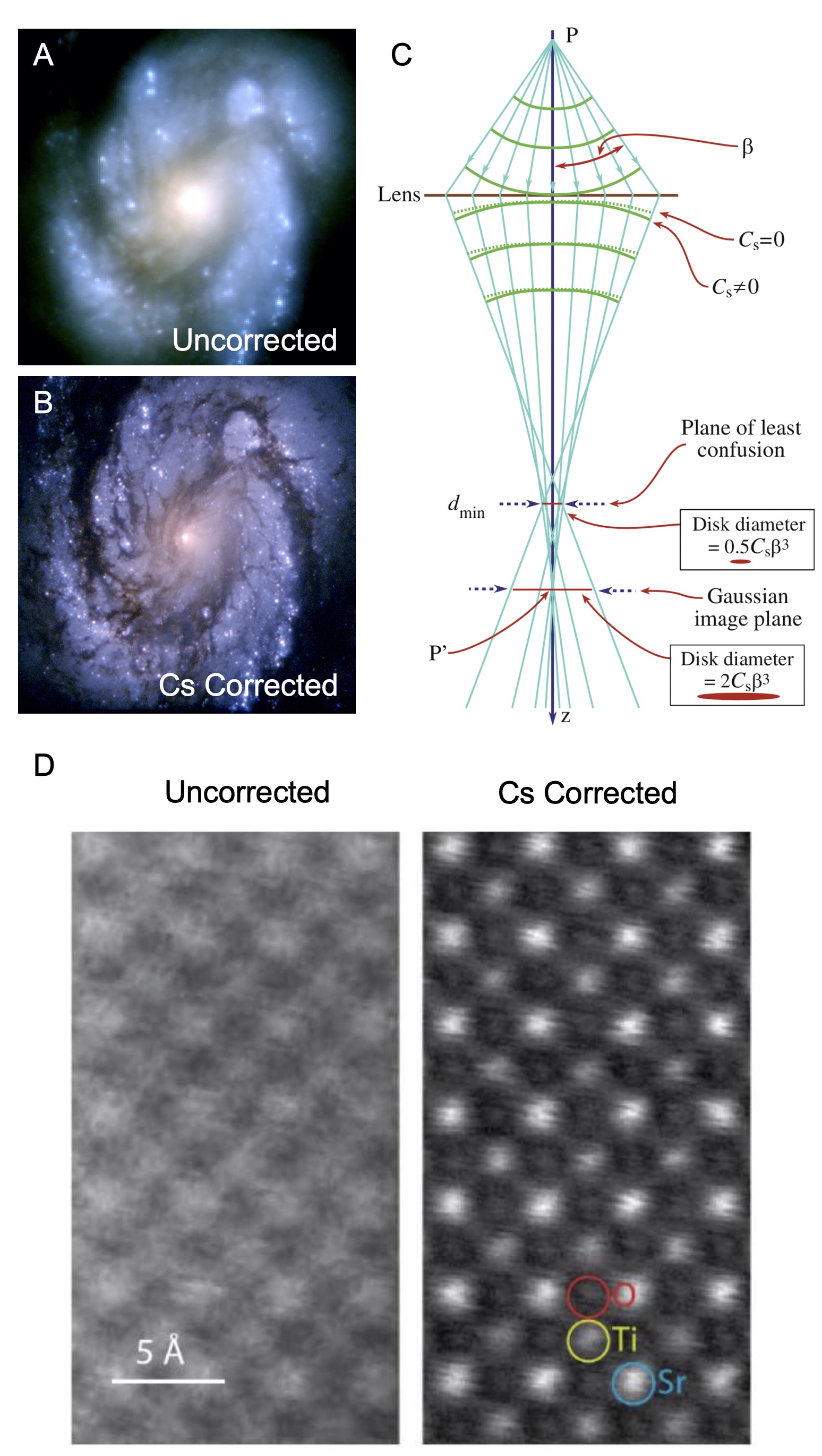}
\caption{Hubble Space Telescope photographs of Spiral Galaxy M100 taken before (A) and after (B) $C_S$ correction. Reproduced with permission from ESA/Hubble Project. (C) Illustration of the effect of spherical aberration on the electron wavefront and the resulting probe quality. Reproduced with permission from Reference \citenum{Williams2009}. (D) Comparison of uncorrected and $C_S$-corrected image of STO. Reproduced with permission from Reference \citenum{Carter2016}.}
\label{spherical_aberration}
\end{figure}

Perhaps no recent technological development has had greater impact on the STEM illumination system than probe aberration-correction. In traditional visible light optics, spherical ($C_S$) aberration can be eliminated through the use of compound and aspheric lens systems, greatly improving spatial resolution. An example of the dramatic effect of this correction is given in Figures \ref{spherical_aberration}.A--B, which show before and after images taken by the Hubble Space Telescope during an upgrade of its mirror. In contrast, pioneering work by Scherzer in 1936 showed that these aberrations cannot be eliminated from rotationally-symmetric electron lenses, leading to ``blurring'' of the electron probe and degrading image quality, as illustrated in Figure \ref{spherical_aberration}.C.\cite{Scherzer1936} However, a decade later in 1947 the same author described various ways to correct these distortions by relaxing the constraints imposed in his earlier paper.\cite{Scherzer1947} In subsequent years it was recognized that a multipole lens system, consisting of quadropole, octopole, and later sextupole components, would be needed to relax the rotational symmetry constraint, but the implementation of this system was hampered by mechanical and electronic instabilities, as well as a lack of computing power to automate the correction process.\cite{Hawkes2009} The community pursued many different designs, with Rose's $C_S$ correction proposal\cite{Rose1981} in 1981 ultimately leading to a practical implementation\cite{Haider1998} by Haider, Urban, and others in 1998. This development coincided with the development of fast computers to adjust the correction on the fly and minimize residual aberrations, as designed by Krivanek \textit{et al}.\cite{Krivanek1997} and eventually implemented in the first commercially available instrument by Batson \textit{et al.} in 2002.\cite{Batson2002} The importance of this new capability is twofold: $C_S$ correction both increases point-to-point spatial resolution by minimizing the curvature of the electron wavefront, as shown in Figure \ref{spherical_aberration}.D, and enables the use of larger condenser apertures, allowing for more current to be packed into a smaller probe. This latter benefit is particularly important, since X-ray fluorescence (for EDS) and inelastically scattered electron (for EELS) signals are proportional to electron current density. With improved signal-to-noise, as well as better detectors (described in Section \ref{section:detectors}), the first sub-nanometer\cite{Muller1993, Browning1993} and finally atomic-scale\cite{Muller2008b} EELS mapping was performed, completing a richly defined and multifaceted picture of the atomic world. In their seminal work,\cite{Muller2008b} Muller \textit{et al.} reported direct spectroscopic measurement of atomic sublattices in a La$_{0.7}$Sr$_{0.3}$MnO$_3$ (LSMO) / SrTiO$_3$ (STO) multilayer sample. They presented spectra for the Ti $L_{2,3}$ ionization edge, revealing distinct changes in local bonding at the interface. This new capability helped inform our understanding of the competition between electronic and chemical reconstructions,\cite{Chambers2010,Nakagawa2006,Warusawithana2013} as well as the behavior of multiferroics\cite{Valencia2011a, Yadav2016, Spurgeon2014} and many other oxides.

\begin{figure}
\includegraphics[width=0.7\textwidth]{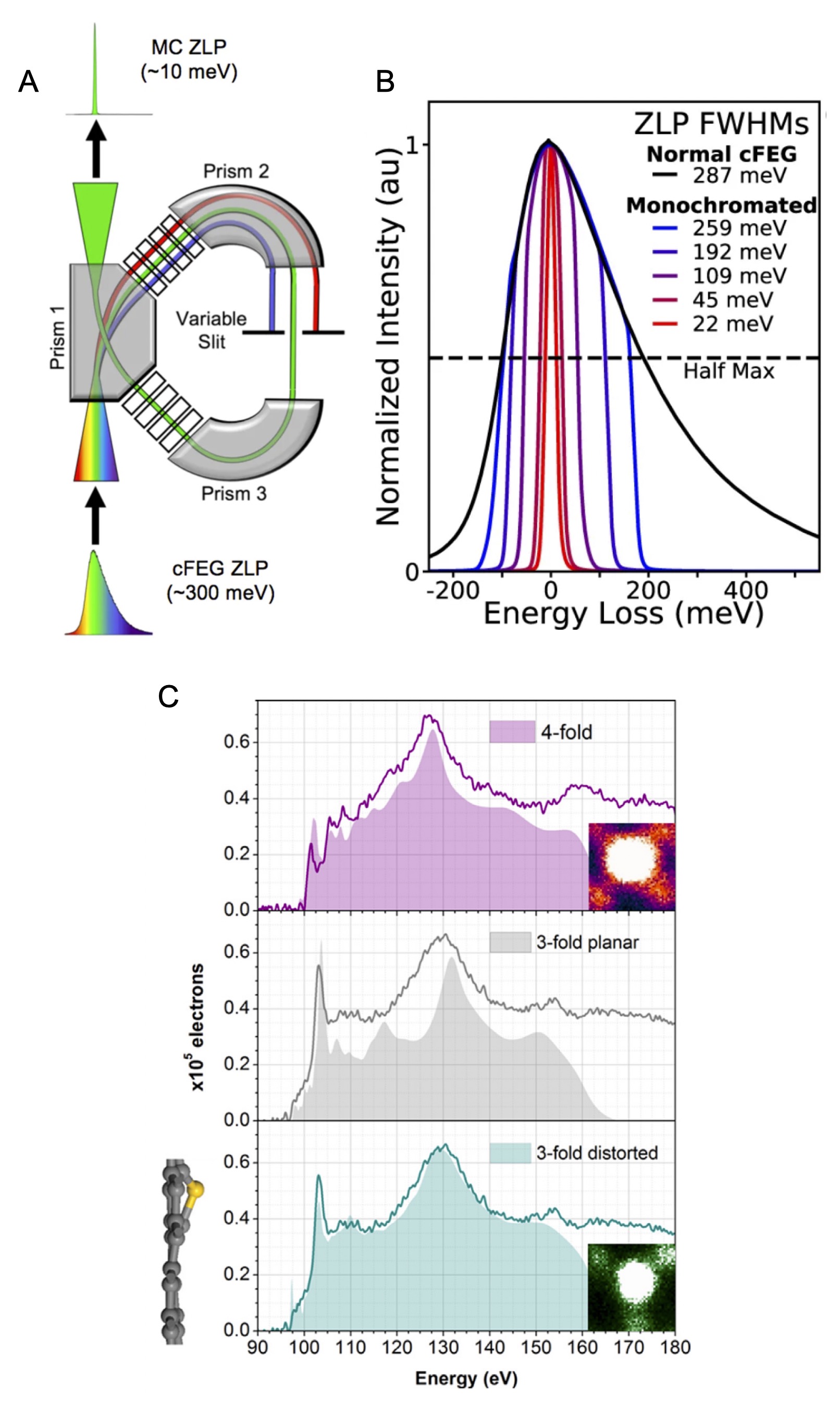}
\caption{(A--B) Schematic of the Wien-type monochromator and associated changes in the energy spread of the electron zero loss peak. Reproduced with permission from Reference \citenum{Hachtel2018}. (C) Monochromated low-loss EELS spectra for different configurations of a substitutional impurity in Si. Reproduced with permission from Reference \citenum{Ramasse2017}.}
\label{monochromator}
\end{figure}

Beyond improvements in spatial resolution, microscope designers have also pursued routes to improve the energy resolution of the electron beam for spectroscopy. As mentioned earlier, the choice of electron source is one of the important factors in determining its energy spread; for a cold FEG this spread is $\sim0.3$ eV, but with large asymmetric zero loss peak tails that complicate low loss measurements.\cite{Erni2005} Early on, a Wien-type monochromator was developed by Boersch \textit{et al.}, which effectively selected a narrow energy window by filtering out electrons emitted from the source.\cite{Boersch1964} This design greatly improved the energy spread, but with a drastic reduction in beam current, as shown in Figures \ref{monochromator}.A--B. Many iterations of monochromators have since been developed,\cite{Hachtel2018} each offering increasingly better energy resolution. Many of these designs have been pioneered by the Nion company, founded by Ondrej Krivanek, and exemplified in the series of UltraSTEM microscopes installed at Oak Ridge National Laboratory (ORNL) in the United States and the SuperSTEM Laboratory in the United Kingdom.\cite{Ramasse2017} The latest Nion UltraSTEM 100MC ``HERMES'' instrument can routinely attain sub 20 meV energy resolution, unlocking unprecedented low loss and valence EELS measurements, as shown in Figure \ref{monochromator}.C. This accomplishment is due in no small part to the incredible mechanical and electrical design of these instruments, which permit longer mapping needed to acquire sufficient signal-to-noise using monochromated beams at $\sim20 \times$ reduced current. Furthermore, these new instruments offer both a reduction in the width of the zero loss peak as well as its beam tails, which permits access to more detailed information in the valence band region. For example, Chambers \textit{et al.}\cite{Chambers2018} have investigated the STO using valence EELS, observing the presence of in-gap defect states similar to those obtained by X-ray photoelectron spectroscopy (XPS), albeit at much higher direct spatial resolution.\cite{Chambers2018} Monochromation also improves the sensitivity of core loss spectroscopy and its integration with theory,\cite{Gazquez2016, MacLaren2014} and provides access to vibrational and phonon spectroscopies.\cite{Idrobo2018, Hage2019, Hage2018}

\begin{figure}
\includegraphics[width=0.7\textwidth]{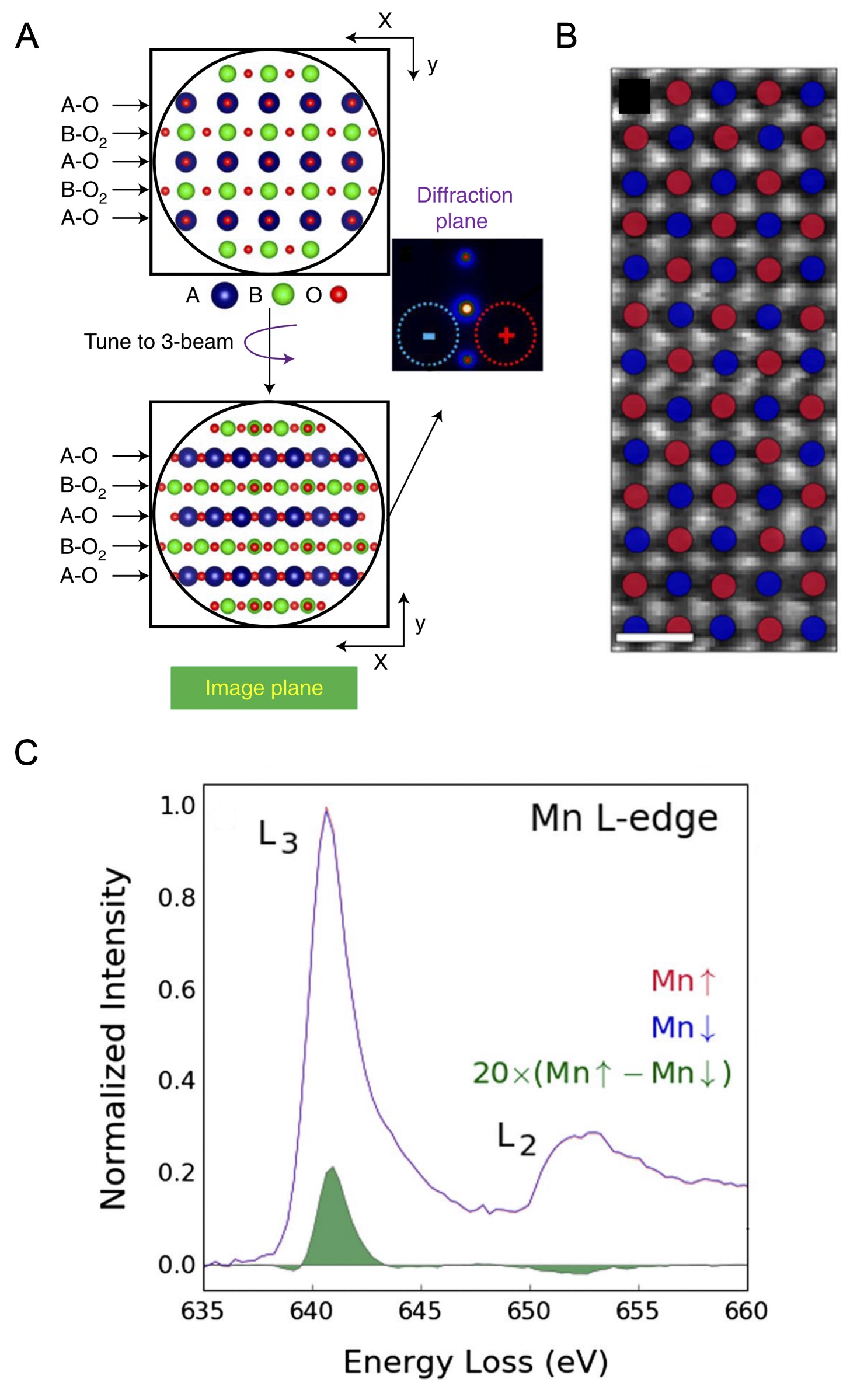}
\caption{(A) Illustration of aberration-corrector tuning to a three-beam condition to select two different chiralities of the lattice, shown in the diffraction plane in red and blue. Reproduced with permission from Reference \citenum{Wang2018}. (B--C) Resulting column-by-column EELS data for the Mn $L_{2,3}$ summed for the two states of the electron beam, illustrating the asymmetry due to local spin polarization. Reproduced with permission from Reference \citenum{Idrobo2016a}.}
\label{emcd}
\end{figure}

As aberration-correction comes into maturity, microscopists have turned their attention to the next series of advances needed to extract deeper insight into the atomic world. Of particular relevance for the oxide community, instrument designers are now focused on controlling the spin state of the electron beam to probe magnetic states of materials. Electron vortex beams, which are formed through the use of micro-machined phase plates, allow for tunable control of orbital angular momentum, albeit not quite yet at atomic-resolution.\cite{Verbeeck2010a, McMorran2011, Uchida2010} Another newly-developed technique called electron magnetic circular dichroism (EMCD) is analogous to X-ray magnetic circular dichorism (XMCD), but offers vastly superior spatial resolution. In this approach, the aberration corrector is used to tune the phase of electron beam and the difference between EELS data collected in two polarization states is used to resolve local changes in spin on the lattice, as shown in Figures \ref{emcd}.A--B.\cite{Wang2018,Rusz2016,Rusz2014,Idrobo2016a} This technique has been successfully applied to double perovskite materials, such as Sr$_2$FeMoO$_6$,\cite{Wang2018} where it was used to determine the degree of spin ordering at the atomic scale. While the measured effect is small for most systems (such as LaMnAsO shown Figure \ref{emcd}.C), theoretical calculations have shown that in principle this method may even be extended to three dimensions\cite{Negi2018} and single atoms.\cite{Negi2019} When coupled with new developments in sample environments, these techniques can further bridge structure, chemistry, and functionality.

\subsection{Sample Environment}

\begin{figure}
\includegraphics[width=\textwidth]{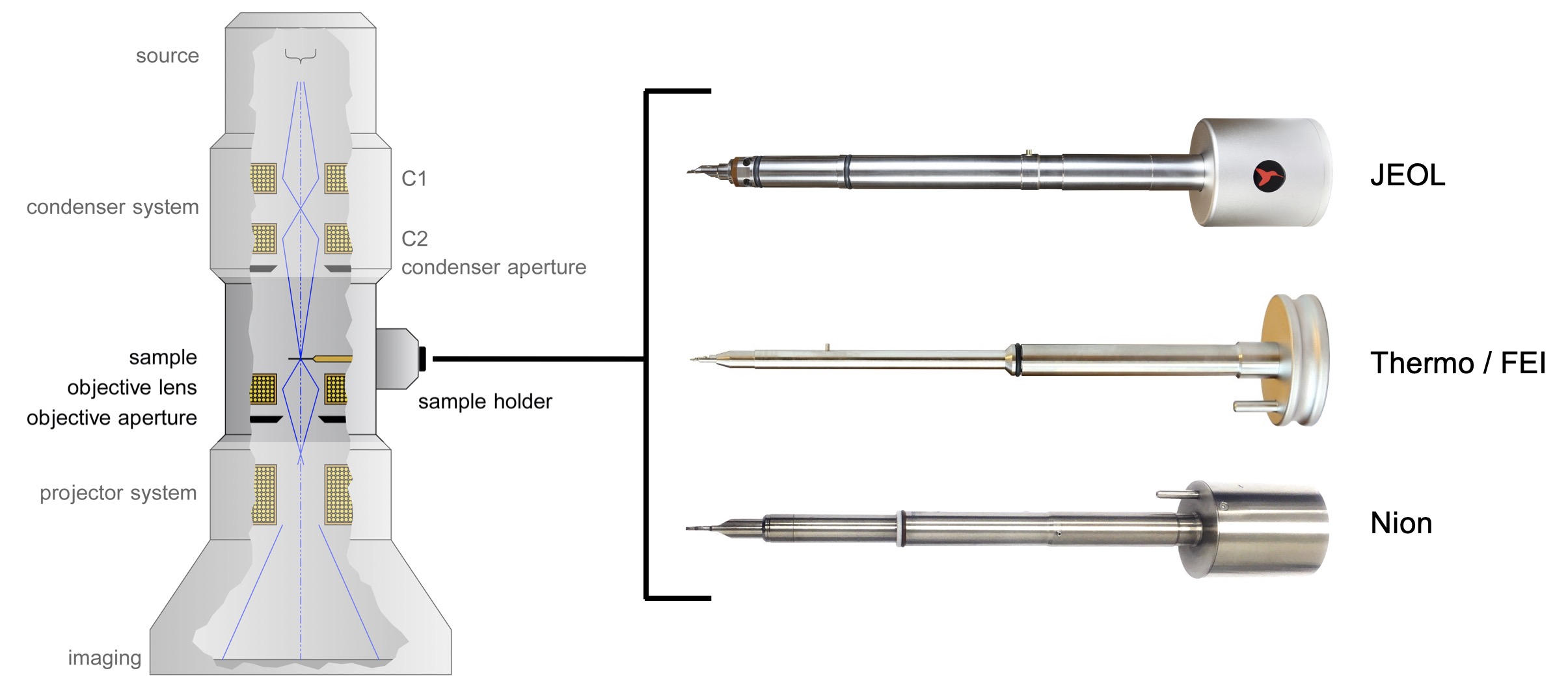}
\caption{Cross-section illustration of the microscope, showing the placement of the sample holder inside the column and a selection of holder designs from different vendors. From top to bottom: MEMS heating-biasing holder for JEOL microscope (Hummingbird Scientific), single-tilt tomography holder for Thermo / FEI microscope (Hummingbird Scientific), and double-tilt side-entry holder for Nion microscope (Nion Company). Illustration adapted with permission of Dr. Eric Jensen. Photographs from Hummingbird Scientific and Nion.}
\label{holders}
\end{figure}

Alongside improvements in the electron source and imaging optics, the design of sample environments has evolved considerably over the past decades. Many of the improvements in spatial and spectroscopic resolution described the last section depend on better control of sample positioning and drift, which have been impressively refined by manufacturers. The sample holder represents a link between the outside world and the high vacuum environment of the microscope. All holders provide the ability to translate sample in X, Y, and Z directions, with some also offering the ability to tilt and rotate. Given the limiting constraints of the electron pole piece, these holders must be intricately designed to accommodate the necessary gears, wiring, vacuum seals, and electronics into the objective gap $< 5$ mm in size. As shown in Figure \ref{holders}, microscope manufacturers have devised ingenious designs to fit the necessary components into such a small form factor. The majority of these are based on a side-entry holder rod design that, while flexible and permitting feedthroughs for \textit{in situ} experimentation, can couple environmental noise into the column (pressure changes, thermal expansion, etc.). Manufacturers have turned to the use of external enclosures to mitigate environmental noise, which can be substantial.\cite{Muller2006a} A small number of microscopes use designs that do not conform to the typical rod geometry---such as the original cartridge design of Nion instruments---which does sacrifice flexibility. Nonetheless, with such designs it is possible to very low drift rates ($<1$ nm/hr) that enable long, stable mapping for EELS and EDS.\cite{Hotz2017} Mundy \textit{et al.} have utilized these tools to great effect to examine superlattices of the multiferroic LuFe$_2$O$_4$ over large areas with high spatial and energy resolution.\cite{Mundy2016c,Mundy2012d} Stability is particularly important in these studies to examine variations across thick periodic structures, rather than at single interfaces.

\begin{figure}
\includegraphics[width=0.55\textwidth]{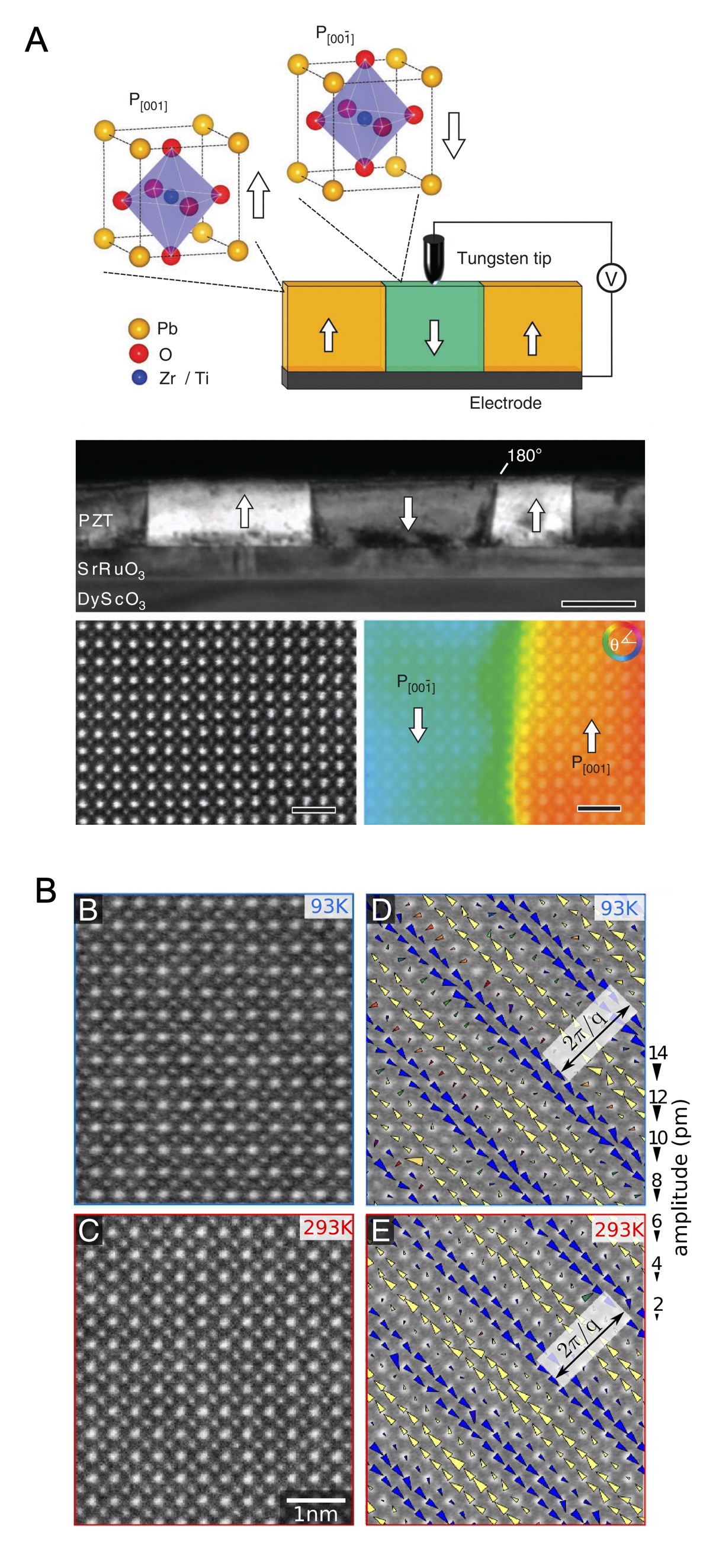}
\caption{(A) \textit{In situ} ferroelectric switching in PZT using a local electrode (top) and resulting changes in polarization. Reproduced with permission from Reference \citenum{Gao2011}. (B) Visualization of charge ordering phase transition and associated lattice distortions at cryogenic temperatures from HAADF images. Reproduced with permission from Reference \citenum{ElBaggari2018}.}
\label{in-situ}
\end{figure}

At present there is a wide array of stages\cite{Gatan} for \textit{ex situ} analysis, encompassing multiple tilt axes ($\alpha$ and $\beta$), rotation, and low X-ray backgrounds, some of which are shown in Figure \ref{holders}. However, some of the most impactful ongoing developments relate to new \textit{in situ} environments to examine dynamic changes in materials. While many dedicated environmental TEMs have been designed to introduce gaseous\cite{Boyes1997,Gai2002,Sharma2001} and liquid\cite{Gai2002a} environments, it was the development of microelectromechanical system (MEMS)-based platforms that led to an explosion of \textit{in situ} studies by making experiments more flexible and reconfigurable. These mass-produced chips allowed for studies of heating,\cite{Allard2012, Damiano2008} electrical biasing,\cite{Garza2017} liquids,\cite{Ross2015} and mechanical properties,\cite{Haque2003} at progressively higher resolution. There is now a variety of competitors in the arena, each producing a dizzying array of holders and chip designs. \textit{In situ} platforms have helped us better understand the functionality of oxides such as ferroelectrics (FEs).\cite{Catalan2012, Gao2011, Nelson2011} Electrical biasing holders developed in the late 2000s enabled \textit{in situ} local switching of ferroelectric domain structures in PbZr$_{1-x}$Ti$_x$O$_3$ (PZT)\cite{Gao2011} and BaTiO$_3$ (BTO).\cite{Polking2012} As shown in Figure \ref{in-situ}.A, many of these designs used an electrode tip that was brought into contact with the surface of the sample to apply a local electric field, causing changes in domain wall morphology. Alternative designs utilized in plane electrodes to apply more uniform electric fields to systems such as BiFeO$_3$\cite{Winkler2014} and BTO / FeGa.\cite{Brintlinger2010} These methods helped show that pinning by misfit dislocations can impede domain wall nucleation and propagation.\cite{Gao2014,Winkler2014}

The dynamics of oxygen vacancies can also be directly examined in the microscope through the use of heating holders that can access temperatures in excess of $1000~^{\circ}$C. It should be noted that in most of these experiments the sample is exposed to the vacuum of microscope ($p_{O_2} = 10^{-8}$ Pa), which creates a highly reducing environment. Early studies examined the complex distribution of phases that can form in (La,Sr)FeO$_{3-\delta}$ under reduction, including Ruddlesden-Popper (RP), Brownmillerite, and Aurivillius structures.\cite{Klie2002, Klie2002a} EELS measurements allowed for direct examination of local oxygen coordination changes alongside observed structural changes. More recent studies have looked at phase transitions in Li$_x$CoO$_2$ cathode materials,\cite{Sharifi-Asl2017} as well as spin state transitions in Ca$_3$Co$_4$O$_9$.\cite{Yang2009} Still, less work has been done in this area and there are many opportunities to expand on these methods.

While not truly an \textit{in situ} method, since it is not performed inside the microscope, our understanding of oxygen defects and associated phase transitions has also been informed by the use of ionic liquid gating. In this approach, an ionic liquid such as 1-ethyl-3-methylimidazolium hexafluorophosphate\cite{Lang2014} is applied to the surface of a sample, which is then electrically biased. The field is removed and samples are extracted using a standard site-specific lift out process. This approach allows for the application of a much higher carrier density than could ever be achieved with a dielectric due to breakdown.\cite{Ueno2010,Ye2010,Asanuma2010} In this way the dynamics of oxygen vacancy transport have been explored in a systems such as SrCoO$_{3-x}$ / La$_{0.45}$Sr$_{0.55}$MnO$_{3-y}$\cite{Cui2017} and La$_{0.8}$Sr$_{0.2}$MnO$_3$.\cite{Ge2015} Metal-insulator transitions, which otherwise would unattainable due to the required charge density, can also be probed in systems such as STO,\cite{Ueno2010} La$_{0.8}$Ca$_{0.2}$MnO$_3$,\cite{Dhoot2009} VO$_2$,\cite{Nakano2012} and NdNiO$_3$.\cite{Dong2017,Asanuma2010} An alternative to the ionic liquid is the use of resistive switching designs based on local electrodes,\cite{Cooper2017} described above.

One technique that has seen a surge of interest is cryo electron microscopy, which was awarded the Nobel prize in 2017 for its important contributions to biology.\cite{Shen2018} Biological studies depend on preserving hydrated organic samples in the harsh microscope vacuum during intense electron beam exposure, which can cause considerable radiolysis damage.\cite{Egerton2004} The use of specialized holders capable of maintaining stability for high-resolution imaging near liquid nitrogen temperatures ($\sim100$ K), coupled with direct electron detectors (described in Section \ref{section:detectors}), has greatly improved the acquisition of data from low-contrast stained organic samples. In the physical sciences, there has been a long history of using these holders to examine oxides, even approaching liquid He temperatures ($\sim20$ K);\cite{Yaicle2005} however, their spatial resolution has typically been limited by instabilities. Still, low-temperatures phase transformations in manganites such as Pr$_{0.5}$Ca$_{0.5}$Mn$_{0.97}$Ga$_{0.03}$O$_3$ have been studied using electron diffraction,\cite{Yaicle2005} as have spin state transitions in cobaltites such as La$_{1-x}$Sr$_x$CoO$_3$.\cite{Gulec2014} It is also possible to examine charge\cite{ElBaggari2018} and vacancy ordering\cite{Rui2019} transitions at high spatial resolution, as shown in Figure \ref{in-situ}.B. While exotic magnetic states such as skyrmions have also been examined,\cite{Shibata2017,Matsumoto2016} recent work has primarily focused on the use of cooling to mitigate beam damage in sensitive battery materials such as Li oxides.\cite{Li2017a,Li2017b,Lin2017} With the worldwide effort to develop low-temperature systems for quantum information science, there has been renewed focus on cryo holders needed to access quantum phase transitions, such as those associated with topological states ($\sim$mK range).\cite{Lee2019} As of the publication of this chapter, the U.S. Department of Energy and their European and Asian counterparts have begun making substantial investments in this area. Similarly, vendors of \textit{in situ} platforms are launching new designs, with the goal to achieve both very low temperatures and high imaging stability in the coming years.

\subsection{Detectors}\label{section:detectors}

\begin{figure}
\includegraphics[width=0.9\textwidth]{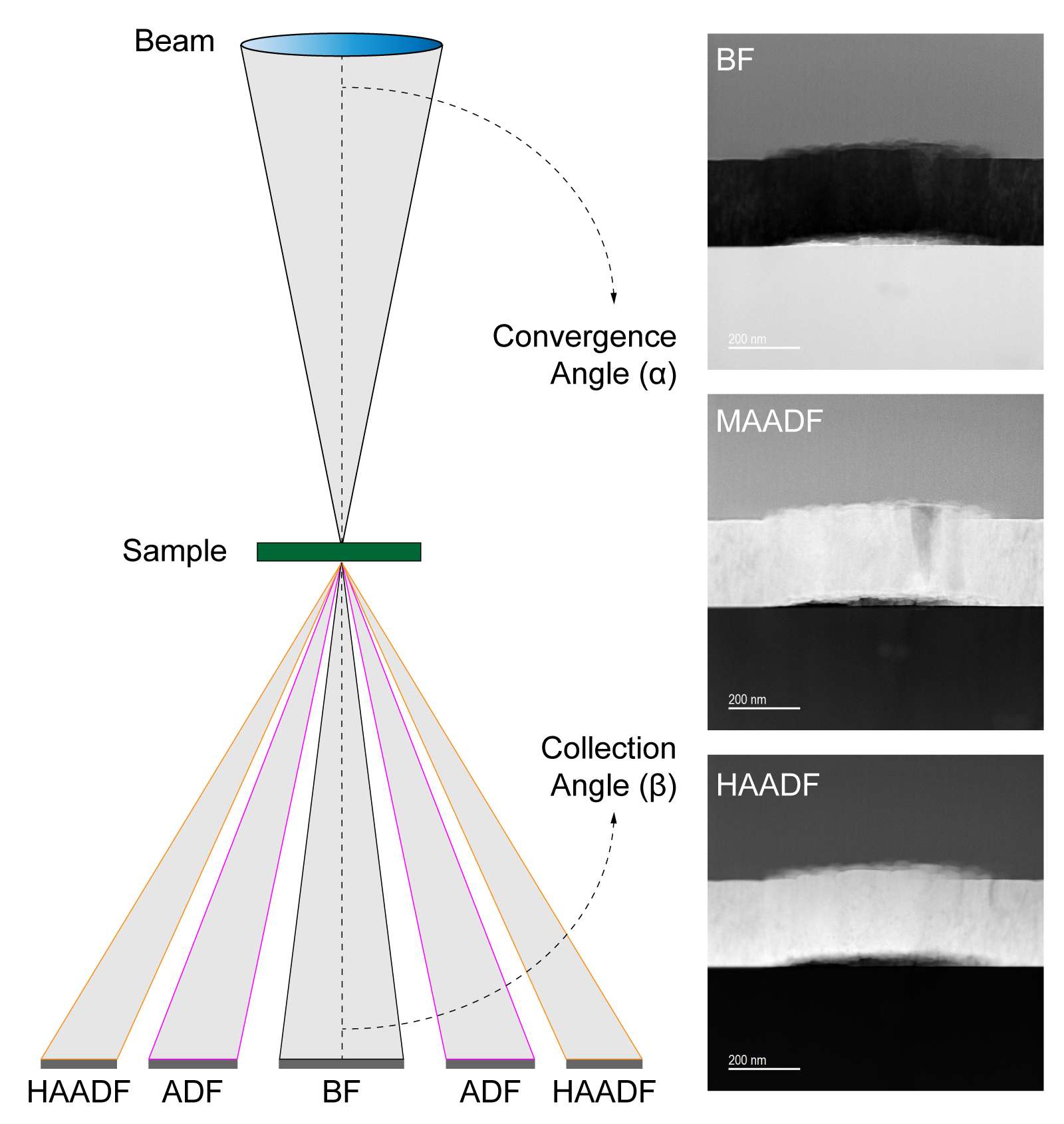}
\caption{Illustration of STEM cross-sectional detector geometry and corresponding images generated from a UO$_2$ thin film sample, showing the variation in contrast mechanisms with collection angle.}
\label{detectors}
\end{figure}

Armed with better illumination and a full range of sample environments, modern instruments are also well-equipped with a suite of detectors capable of recording the diverse signals generated by the electron-sample interaction. In recent years, high-speed imaging cameras, scanning nanodiffraction (today commonly known as 4D-STEM), and improved spectrometers have had a large impact on the study of oxides. Detectors have come a long way since the photographic film used in the early days of microscopy. These first instruments used the same suspension of silver halide solution in gels as in traditional light photography, which came in a number of speeds (different grain sizes) and offered a relatively high detector quantum efficiency (DQE).\cite{Williams2009} In contrast, most STEMs today are equipped with several detectors: a phosphorescent ZnS screen for basic alignment and positioning, semiconductor detectors for coherent/incoherent imaging, and charge coupled detectors (CCDs) for diffraction. The ZnS screen is increasingly hidden in newer instruments, which use video cameras for remote viewing, but its characteristic green glow is still commonplace in microscopy labs. Semiconductor detectors based on $p$-$n$ junctions are robust, highly configurable, and can be easily shaped into the annular geometry used for most STEM imaging modes; however, these detectors have a large dark current and relatively poor DQE at low signal intensity.\cite{Williams2009} As shown in Figure \ref{detectors}, the scattering geometry of these detectors is determined by the convergence and collection semi-angles ($\alpha$ and $\beta$, respectively). By selecting different camera length settings, a series of circular and annular detectors can be arranged to bisect various electron scattering angles, giving rise to different modes of image contrast. The corresponding collection angles are loosely defined as follows: bright-field (BF) $<10$ mrad, low-angle/medium-angle annular dark-field (LAADF/MAADF) $25-60$ mrad, and high-angle annular dark-field (HAADF) $>60$ mrad. Each mode offers different and complementary information about a sample: BF produces phase-contrast images that are sensitive to diffraction effects, while LAADF/MAADF produces images more sensitive to light elements and strain, and HAADF produces images that are proportional to atomic number $Z^{\sim1.7}$ and insensitive to defocus and thickness.\cite{Pennycook2011a} With probe aberration-correction, HAADF imaging has become the de facto imaging mode for oxides, offering both the highest spatial resolution and readily interpretable image contrast (bright = heavy, dark = light atoms), as discussed in Section \ref{sec:illumination}.

\begin{figure}
\includegraphics[width=0.8\textwidth]{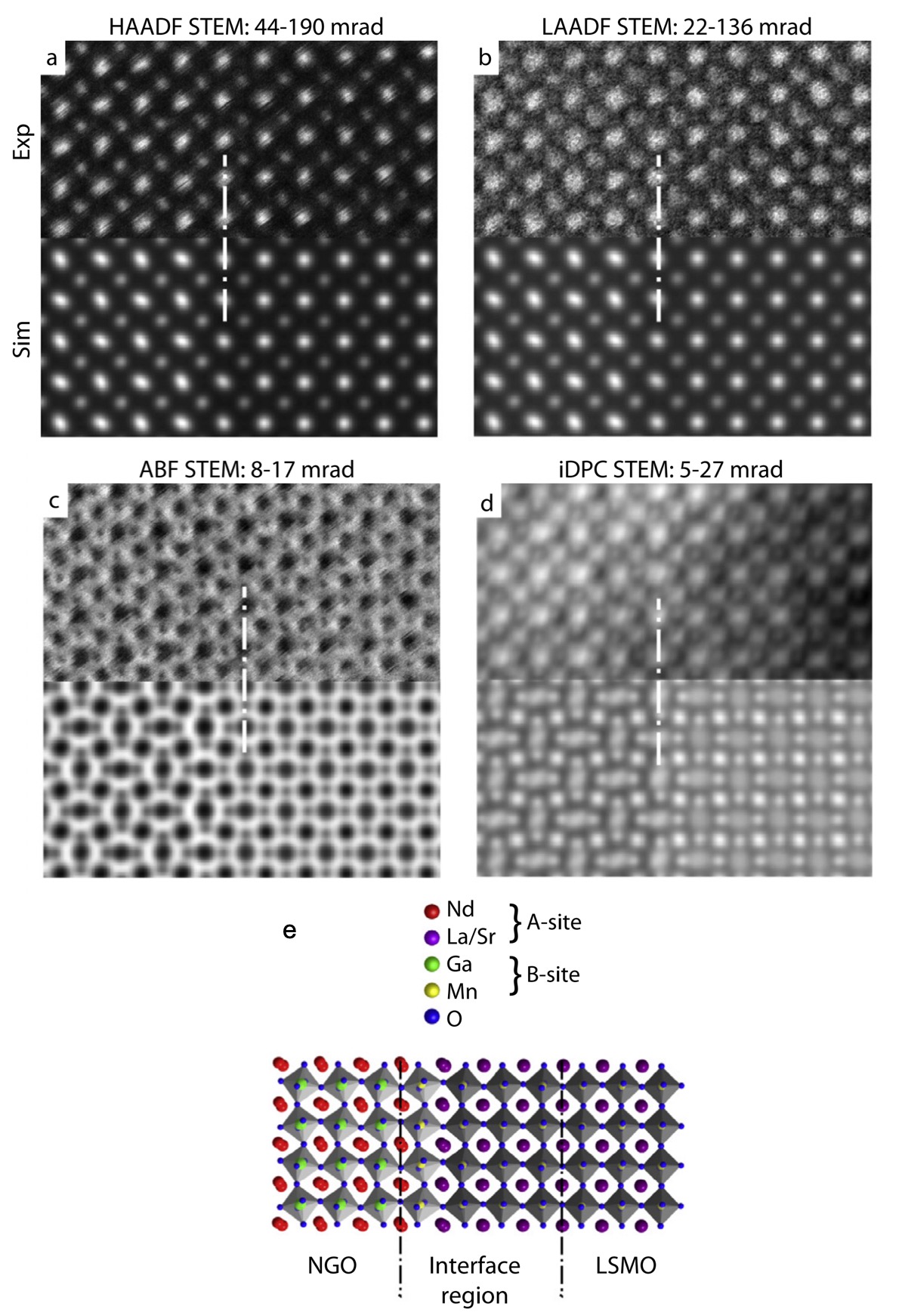}
\caption{Comparison of contrast generated by HAADF, LAADF, ABF, and DPC imaging modes to examine cation and oxygen structures in a NaGdO$_3$ / La$_{1-x}$Sr$_x$MnO$_3$ interface. Reproduced with permission from Reference \citenum{Gauquelin2017}.}
\label{oxygen}
\end{figure}

Of particular concern to studies of oxide interfaces, there are several other modes that can highlight low-contrast or otherwise difficult-to-image features of samples. For example, the ABF mode\cite{Okunishi2012,Findlay2009b} offers improved sensitivity to light elements, such as oxygen\cite{Okunishi2009} and hydrogen,\cite{Ishikawa2011a} by utilizing a combination of beam stop and aperture to select only part of the bright-field disk. This approach has been used to visualize oxygen bond distortions and octahedral rotations at interfaces in systems such as BiFeO$_3$.\cite{Kim2017,He2015,Wang2016} However, in contrast to HAADF imaging, the technique is more thickness and defocus sensitive and must be applied carefully.\cite{Gauquelin2017} Differential phase contrast (DPC) imaging based on segmented detectors\cite{Haider1994} is another emerging mode that allows access to important features and functionality of oxides. This approach forms an image by subtracting signals from different detector quadrants, highlighting deflection of the electron beam due to local electric fields in materials such as BaTiO$_3$.\cite{Shibata2012,Shibata2010} In addition, it is possible to image both light and heavy elements simultaneously by combining signals from different parts of the detector.\cite{Gauquelin2017, Cooper2017} A comparison of four commonly-used imaging modes is given in Figure \ref{oxygen}. Importantly, these detector options represent a new paradigm for electron microscopy. Rather than selecting a particular detector configuration prior to an experiment, information about the scattered signal can be collected and manipulated after the fact to highlight features of interest. This approach has a strong synergy with emerging big data techniques (described in Section \ref{sec:analytics}) that seek to mine large data sets for features and correlations that would otherwise go undetected by human operators.

In addition to the semiconductor detectors used to form STEM images, CCDs are used for imaging in parallel-beam TEM mode and diffraction in STEM mode. These systems offer up to 4k $\times$ 4k pixel resolution with low noise, high dynamic range, and a relatively good DQE of $\sim0.5$.\cite{Williams2009} When binned, they can attain readout rates of several hundred frames per second (fps), which led to their early adoption for high-speed \textit{in situ} imaging. CCDs also serve as the main camera for EELS systems. While this technology is relatively affordable, proven, and commonplace, it faces growing competition from direct electron detectors (DDs). First developed in the early 2000s,\cite{Milazzo2005,Faruqi2003} DDs forego the traditional scintillator and fiber optic coupling that cause lateral charge spread, offering improved pixel resolution and much better DQE for low-dose applications.\cite{Grob2013,Kuijper2015} These features have led to their broad adoption by the biological community for cryo electron microscopy, where low-contrast structures are common.\cite{Wu2016,Bammes2012,Xuong2007} Recently, attention has turned to the use of DDs as a spectrometer camera for EELS,\cite{Hart2017} where their improved dynamic range and high detection efficiency enable collection of a wide range of energy losses.\cite{Cheng2019,Maigne2018} The reduced point spread function of DDs effectively allows for a wider spectral field of view without sacrificing energy resolution, which is important for measurements of systems containing a range of alloying elements with well-separated ionization edges. Very high energy EELS is now a real possibility, approaching transition energies previously only accessible to synchrotron-based X-ray techniques. For instance, Maclaren \textit{et al.} have shown that it is possible to measure transition metal $K$ shell transitions (5--10 keV energy loss) with excellent resolution.\cite{MacLaren2018}

\begin{figure}
\includegraphics[width=0.45\textwidth]{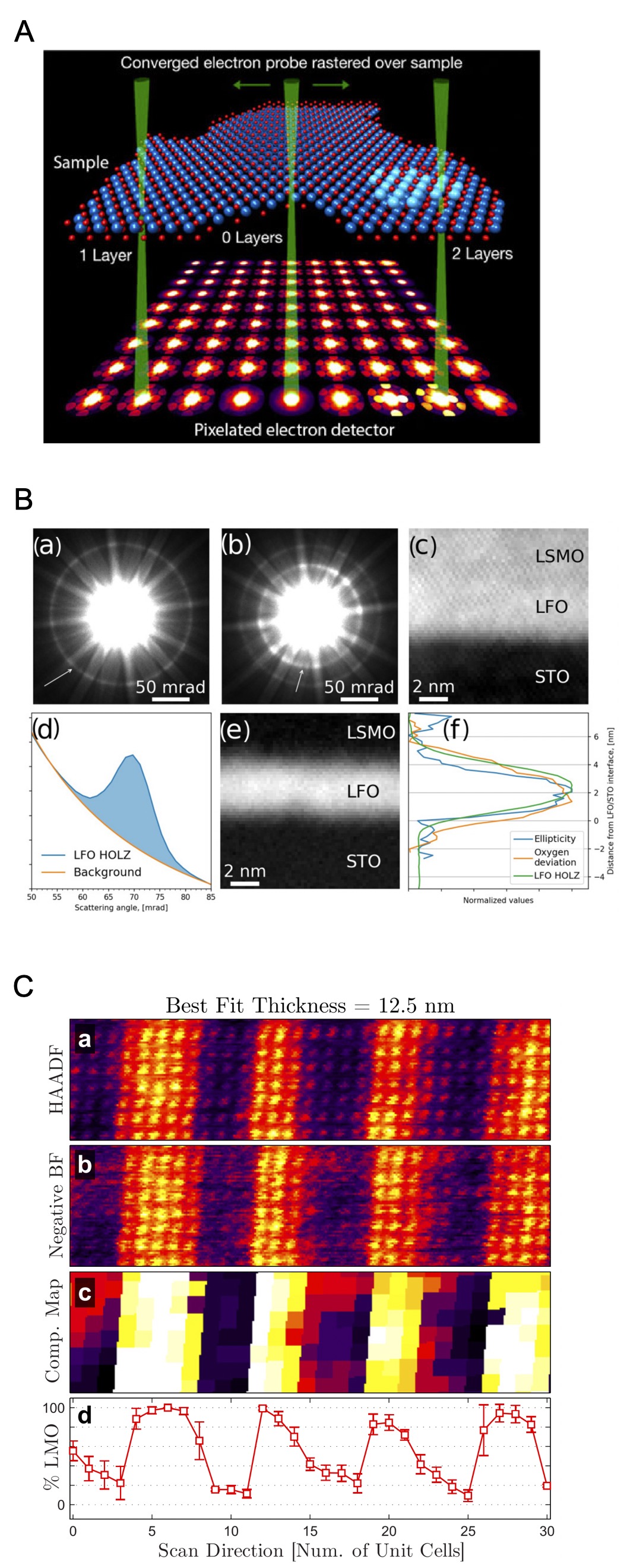}
\caption{(A) Illustration of 4D-STEM acquisition. Reproduced with permission from Reference \citenum{Ophus2019}. (B) 4D-STEM measurement of octahedral rotation in LSMO / LFO. Reproduced with permission from Reference \citenum{Nord2019}. (C) Determination of unit-cell level composition in LSMO / STO by fitting simulations to 4D-STEM data. Reproduced with permission from Reference \citenum{Ophus2017b}.}
\label{4dstem}
\end{figure}

Beyond improvements in conventional imaging and spectroscopy, high-speed detector technologies have unlocked 4D-STEM diffraction that leverages today's vast computational and data storage capabilities. STEM diffraction has long been used to measure beam aberrations,\cite{Cowley1978, Cowley1989} examine nanoscale strain,\cite{Clement2004,Ozdol2015} and probe local order\cite{Hirata2011} under names such as convergent beam electron diffraction (CBED) and nanobeam electron diffraction.\cite{Ophus2019} Recently, technologies based on active pixel sensors (APS)\cite{Milazzo2005,Mendis1997,Dierickx} and pixel array detection (PAD)\cite{Ansari1989,Caswell2009,Ercan2006} have yielded the high sensitivity and fast readout speed needed to collect entire diffraction patterns during scanning. Now, rather than simply acquiring images using the aforementioned fixed detectors (ADF, BF, etc.), almost the entire scattered signal can be collected at each point of a sample, as shown in Figure \ref{4dstem}.A. In practice, it is easy for even modest pixel samplings and diffraction pattern resolutions to grow file sizes on the order of hundreds of GB or even TB in a few minutes. However, having access to the entire scattered signal means that decisions on what imaging mode to use must no longer be made directly at the microscope. Rather, 4D-STEM datasets can be reviewed and analyzed after the fact both manually and by automated data processing routines. This shift to collecting all the data available represents the future of microscopy and materials science in general. The value of this approach for oxides lies in the fact that the diffracted intensity contains information about crystallinity, local order, strain, thickness, electric fields, and much more.\cite{Ophus2019} For example, fitting of CBED patterns to simulations allows for precise determination of octahedral rotations in LaFeO$_3$ (see Figure \ref{4dstem}.B),\cite{Nord2019} local composition in La$_{0.7}$Sr$_{0.3}$MnO$_3$ / STO,\cite{Ophus2017b} (see Figure \ref{4dstem}.C), and ferroelectric polarization in BaTiO$_3$.\cite{LeBeau2011a} 4D-STEM can also provide an alternative to the aforementioned DPC measurements, which require dedicated, fixed detectors. Rather, by measuring the center of mass of the diffraction pattern\cite{Muller-Caspary2018,Muller-Caspary2017} it is possible to detect small displacements of the STEM probe due to electric fields, uncovering octahedral distortions in DyScO$_3$,\cite{Hachtel2018a} ferroelectric domains in BiFeO$_3$,\cite{Tate2016} and electric field variations in STO.\cite{Muller2014} As this technique continues to mature, new ways to process and handle the large volumes of diffraction data are starting to take shape, such as those based on machine learning.\cite{Li2019,Martineau2019} Many exciting developments are expected in the coming years.

\subsection{Data Analytics}\label{sec:analytics}

As the preceding sections have outlined, modern microscopes can quickly produce rich, high-resolution structural and chemical information. However, data alone without sufficient context and interpretation is not of much value. Microscopy has long since passed the point of data saturation, where even a beginning user can generate 10--100 GB of images, diffraction patterns, and spectra in a single session---far more than could ever be manually analyzed. Distilling this data down to its salient features is a grand challenge that will change the way materials are conceptualized, synthesized, characterized, and modeled.\cite{Belianinov2015} Alongside instrumentation developments, data science and machine learning are poised to revolutionize the analysis of STEM data due to their ability to classify features in complex, multidimensional data sets with little-to-no human intervention.\cite{Voyles2016} These methods have grown in tandem with our ability to more quickly and accurately simulate microscope data to connect experiments to atomistic models.

At a basic level, it is well known that the signal-to-noise ratio of microscope data can be improved through denoising approaches based on the removal of Poisson noise; the most common examples of this are variants of weighted and non-local principal component analysis (PCA),\cite{Kotula2006, Kotula2003} though the physical interpretation of resulting components can be challenging.\cite{Cueva2012} Beyond merely separating signal components from noise, these dimensionality-reducing methods offer the possibility to highlight correlations in multidimensional imaging\cite{Belianinov2015} and diffraction\cite{Jesse2016} data sets. Sliding fast Fourier transforms (FFTs),\cite{Vasudevan2015} coupled with PCA, have been shown to effectively separate parts of STEM images to detect interfaces or different materials phases.\cite{Vasudevan2016} Clustering analysis methods, such as $k$-means, are also able to extract features in a semi-supervised fashion,\cite{Belianinov2015} though often manual cluster size determination is often needed.

\begin{figure}
\includegraphics[width=0.8\textwidth]{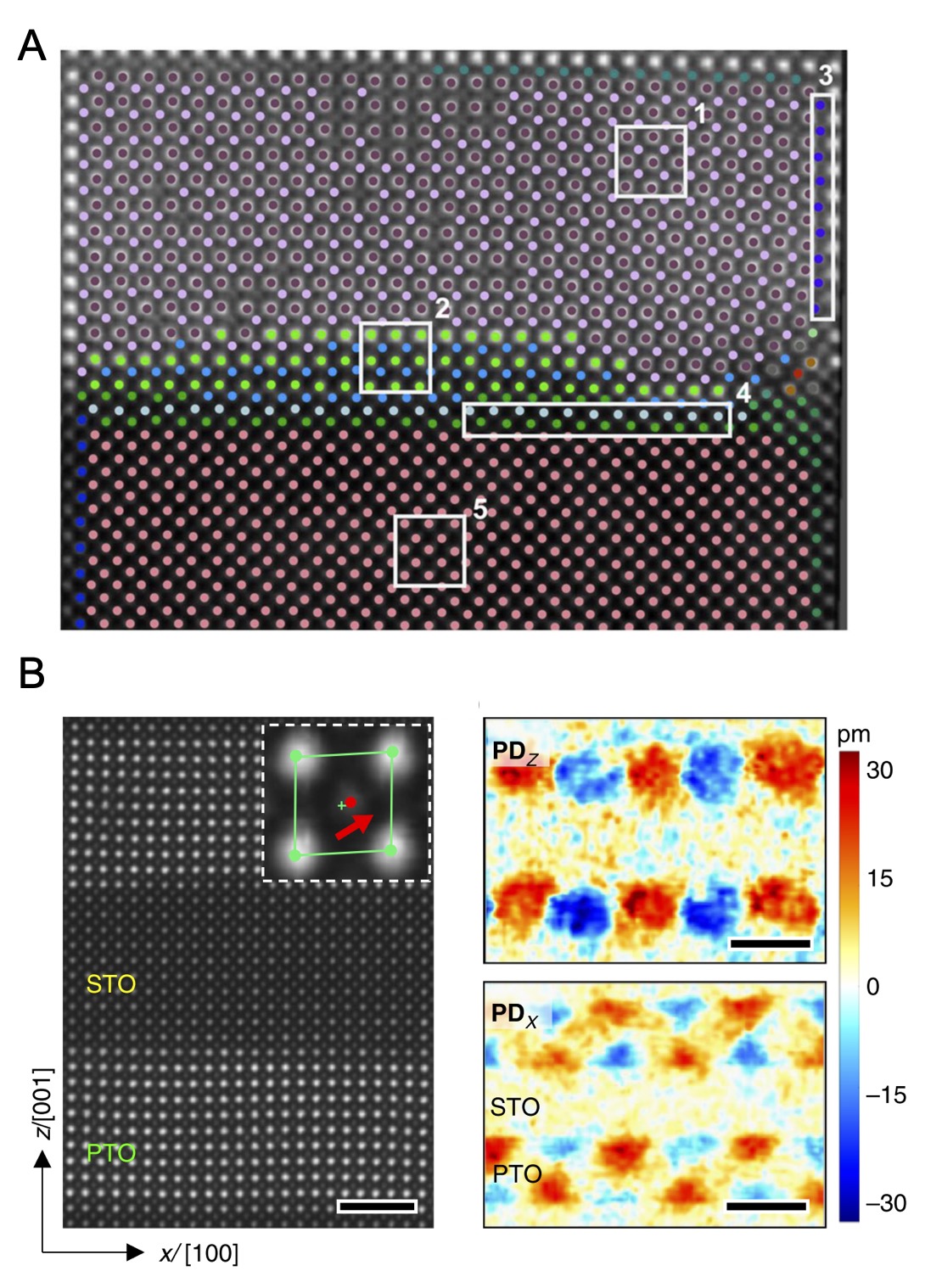}
\caption{(A) Atomic-scale classification of interface states in LaCoO$_3$ / STO using a multi-scale extraction approach. Reproduced with permission from Reference \citenum{Laanait2016}. (B) Quantification of polar vortices in PbTiO$_3$ / STO using principal component analysis to match to phase field simulations to HAADF images. Reproduced with permission from Reference \citenum{Li2017}.}
\label{ml}
\end{figure}

Some of the most exciting developments have centered on the use of machine learning, and, specifically, deep learning, to derive materials properties and functionality from imaging data.\cite{Vasudevan2019, Voyles2016, Kalinin2015} An early study of InGaAs / AlGaAs showed how image simulations may be used to train a neural network to detect structural motifs in high-resolution transmission electron micrographs.\cite{Kirschner2000} More recent examples include the use of CBED simulations to train convolutional neural networks to extract defect configurations from diffraction data more accurately than trial-and-error fitting methods.\cite{Xu2018a,Pennington2015,Pennington2014} Xu \textit{et al.} have applied this approach to rapidly assess orientation and sample tilt in STO.\cite{Xu2018a} Recently, several frameworks have been proposed to extract atom positions and physical descriptors from STEM and other scanning probe data sets.\cite{Kannan2018, Madsen2018, Ziatdinov2017, Vlcek2017} To date, most past work in this space has focused on two-dimensional materials with high-contrast defects relative to their local environment. However, relevant recent examples include reconstruction of octahedral distortions in CaTiO$_3$,\cite{Laanait2019} quantification of polar vortices in PbTiO$_3$ / STO,\cite{Li2017} and identification of interface configurations in LaCoO$_3$ / STO,\cite{Laanait2016} as shown in Figure \ref{ml}.

The emergence of machine learning techniques stands to benefit greatly from high-throughput image simulations that can inform features present in multi-modal STEM data sets. The ability to acquire simultaneous structural and chemical information at the atomic-scale carries with it the tacit promise to resolve underlying defects governing functional properties. However, recovering full information about an object from images or spectra is difficult because of strong electron beam interactions with the sample that introduce artifacts,\cite{Wu2017} as well as our inability to efficiently detect and interpret correlations across large data sets.\cite{Voyles2016,Belianinov2015} The large convergence angle of aberration-corrected electron probes leads to signal broadening as the beam propagates through the sample. The periodicity of the lattice also introduces complex channeling effects that obscure the underlying structure.\cite{Spurgeon2019, Oxley2007, MacArthur2017, Kothleitner2014, Wang2008, Lugg2012, Chen2016, Lugg2014, Dycus2015, Dwyer2008}

These effects are even more pronounced in spectroscopy of interfaces, where local lattice symmetry and chemistry changes modulate channeling in a complex, difficult-to-predict manner.\cite{Allen2017, Spurgeon2016b} Reconstructing the chemical profile across an interface becomes very challenging, even in the case of relatively thin samples in which signal delocalization effects should be minimized. For example, Spurgeon \textit{et al.}\cite{Spurgeon2016b} considered La$_{0.88}$Sr$_{0.12}$CrO$_3$ / Nb:STO interfaces using ionization simulations and found that, even in the case of a perfectly abrupt boundary, delocalization can lead to artificial blurring of the measured chemical profile. Moreover, they observed that the interface profile can vary with the choice of ionization edge, being sharper for more localized, high-energy $K$ shell transitions than for more delocalized $L$ shell transitions. These results show that unless exceptionally thin samples are used,\cite{Lu2017,Lu2013,Lu2014,Lu2018} beam channeling can lead to erroneous conclusions, necessitating the use of image simulations for accurate quantification.\cite{Spurgeon2016b, Allen2012}

Computerized STEM simulations have a long history reaching back to the 1970s and 80s,\cite{Kirkland1987, Ishizuka1977} though the theoretical underpinnings were established in the 50s.\cite{Cowley1957a} The two main image simulation techniques are Bloch wave and multislice calculations, also termed real-space and reciprocal-space image simulations, respectively.\cite{Williams2009} In the Bloch wave approach, we consider that, while many diffracted beams are formed upon interaction of an electron beam with a highly symmetric crystal, only a small number of Bloch waves give rise to the actual image.\cite{Kambe1982, Fujimoto1978} Mathematically, we can describe the propagation of these waves as, 

\begin{equation}
\frac{\partial \psi (\overrightarrow{r})}{\partial z} = \frac{i \lambda}{4\pi} {\nabla_{xy}}^2 \psi (\overrightarrow{r}) + i \sigma V (\overrightarrow{r})\psi(\overrightarrow{r})
\label{eqtn_bloch}
\end{equation}

\noindent where $\lambda$ is the relativistic electron wavelength, $\nabla_{xy}^2$ is the 2D Laplacian operator, $\sigma$ is the relativistic beam interaction constant, and $V(\overrightarrow{r})$ is the electrostatic potential of the sample.\cite{Ophus2017,Kirkland2010} In this method we calculate a basis set assumed to be periodic in all directions that consists of the eigendecomposition of a set of linear equations that approximate the propagation up to a chosen maximum scattering vector ($|q_{max}|$). This maximum scattering vector essentially bounds the accuracy and computation time of the simulation. Next, weighting coefficients are calculated for each element of the Bloch wave basis set, corresponding to different STEM probe positions on the sample, and these coefficients are multiplied by the basis set to determine the final exit wave after the sample.\cite{Ophus2017} This method is computationally efficient, particularly for small simulations, but it is limited in that we only consider diffracted wave propagation in the forward direction of scattering.

\begin{figure}
\includegraphics[width=0.4\textwidth]{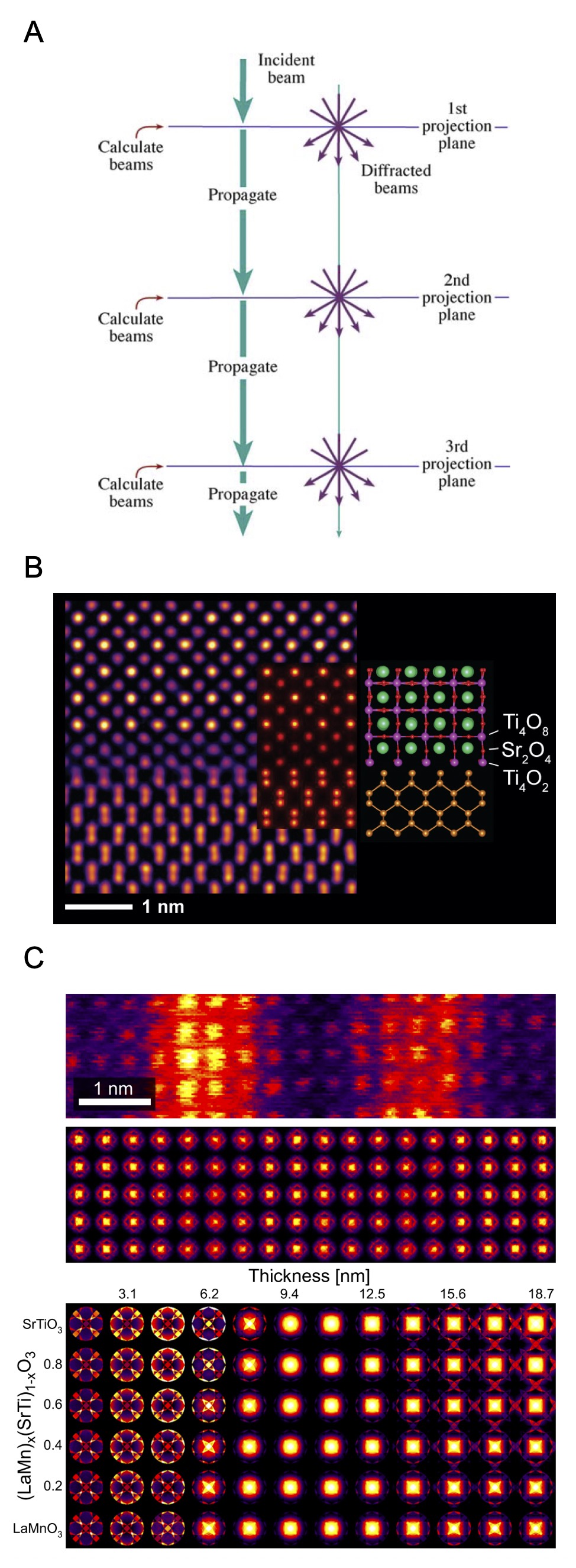}
\caption{(A) Illustration of the multislice algorithm for image simulation. Reproduced with permission from Reference \citenum{Williams2009}. (B) HAADF image of the STO / Ge interface, inset with multislice simulation based on \textit{ab initio} models. Reproduced with permission from Reference \citenum{Du2018}. (C) Quantification of atomic-scale PACBED data based on large scale diffraction simulations in LSMO / STO. Reproduced with permission from Reference \citenum{Ophus2017}.}
\label{multislice}
\end{figure}

The alternative to Bloch wave calculations is the more commonly used multislice (reciprocal-space) approach.\cite{Cowley1957a} While more computationally expensive, this method is generally more accurate, since we calculate all the diffracted beams generated by a point scatterer in the crystal.\cite{Williams2009} As its name implies, the multislice method subdivides the crystal lattice into a number of projection planes $t$, each of which acts as a diffraction grating. As shown in Figure \ref{multislice}.A, the incident beam is propagated through such a plane and all the diffracted beams are calculated. These beams are then passed on through free space to the next projection plane, and the calculation is repeated until the number of planes equals the thickness of the crystal. Mathematically, the left and right hand sides of Equation \ref{eqtn_bloch} are solved separately and calculated for all slices as,

\begin{equation}
\psi_{p+1}(\overrightarrow{r}) = \mathcal{F}^{-1}\{\mathcal{F}\{\psi_p (\overrightarrow{r})e^{i\sigma V_p^{2D}(\overrightarrow{r})} \} e^{-i\pi|\overrightarrow{q}|^2t} \}
\end{equation}

\noindent where the term $e^{-i\pi|\overrightarrow{q}|^2t}$ comes from the Fresnel propagation operator and the term $\psi_p (\overrightarrow{r})e^{i\sigma V_p^{2D}(\overrightarrow{r})}$ relates to the integrated potential in the slice along the beam direction.\cite{Ophus2017} Traditionally the multislice method has been inefficient for large simulations, since the probe propagation must be calculated separately for each probe position. Recently, new simulation programs\cite{Allen2015,Ophus2017} based on graphics processing unit (GPU)-accelerated computing have enabled rapid, highly accurate imaging, diffraction, and spectral simulations based on atomistic models. These simulations are (1) able to scan a wide microscope and sample parameter space to distinguish imaging artifacts from true features, and (2) permit fast matching of images to theoretical structures (such as those calculated by \textit{ab initio} methods). An example of these new codes is the PRISM algorithm,\cite{Pryor2017, Ophus2017} which implements certain approximations of the full multislice routine. PRISM uses an interpolation factor ($f$) that reduces the number of calculated plane waves by a factor $f^2$ at the expense of some numerical accuracy. In practice, it is found that for low interpolation factors the error compared to full multislice is $\sim1\%$, but computation time is reduced by a factor of $f^4$. This approach permits many simulations to be completed more quickly, providing more candidates for matching to experiments as well as helping to generate large training data sets for machine learning. Several examples of the correlation between experiment and simulation are shown in Figures \ref{multislice}.B--C for STO / Ge\cite{Du2018} and La$_{0.7}$Sr$_{0.3}$MnO$_3$ / STO\cite{Ophus2017b} interfaces, respectively. In the former example, Du \textit{et al.} determined a number of possible interface configurations using \textit{ab initio} calculations, whose energy was minimized, ranked, and compared to image simulations for likely experimental sample thicknesses. The most closely matched structures were then refined in light of the experiment and the resulting simulations were used to calculate the band structure of the interface. This approach allows for a rich set of electronic and magnetic properties to be extracted from a local picture of the interface.

\section{Applications}

The technological developments in electron microscopy described in the preceding sections have had a transformative impact on the study of complex oxides and their interfaces. While other area- and volume-average techniques, such as those based on X-ray diffraction and spectroscopy, provide valuable information about these materials, STEM techniques offer a wealth of \textit{directly resolved} structural, chemical, and defect information at the highest spatial resolutions. We next discuss how microscopy has informed our understanding in the areas of ferroelectrics and multiferroics, magnetoelectric heterostructures, and synthesis pathways in complex oxides.

\subsection{Ferroelectrics and Multiferroics}\label{sec:ferroelectrics}

STEM techniques have had a tremendous impact on the development of ferroelectrics and multiferroics. These materials exhibit spontaneous electrical polarization and coupling to other kinds of ferroic order (e.g. ferromagnetism) and are among the most prolific oxides. Bulk ferroelectric ceramics have been studied since World War II, which motivated the design of functional materials for the war effort.\cite{Haertling1999} Subsequent decades saw the development of thin film ferroelectrics for computer memory, radio frequency and microwave devices, and sensors.\cite{Martin2017} Continued refinement of synthesis methods helped achieve precise, atomic-level control of these systems in the 2000s, enabling new properties through interface engineering.\cite{Martin2010} At the same time, aberration-corrected STEM enabled characterization of structure, defects, and chemistry down to the picometer level.\cite{MacLaren2014} It is now possible to routinely design materials interfaces to access novel functionalities not found in nature.\cite{Zubko2011, Rondinelli2011, Eerenstein2006}

\begin{figure}
\includegraphics[width=0.7\textwidth]{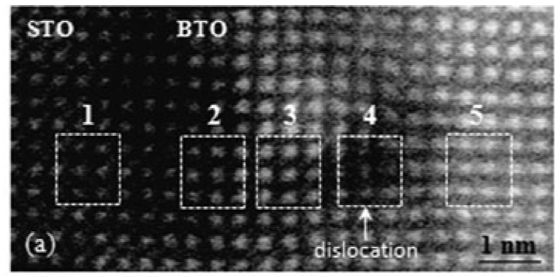}
\caption{HAADF image of dislocation cores in BaTiO$_3$. Reproduced with permission from Reference \citenum{KURATA2009}.}
\label{bto_pto}
\end{figure}

\begin{figure}
\includegraphics[width=0.7\textwidth]{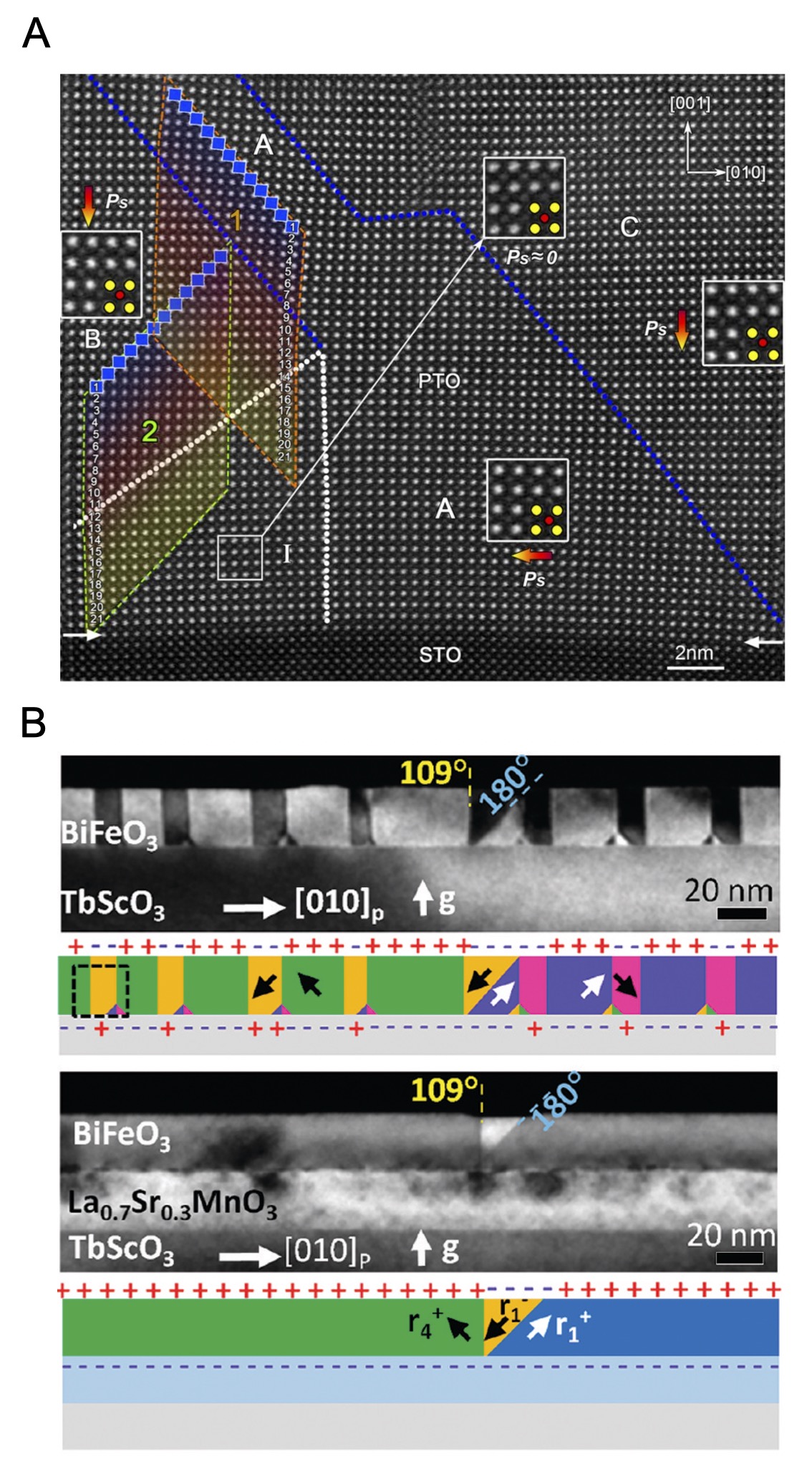}
\caption{(A) HAADF image of domain structure in PbTiO$_3$. Reproduced with permission from Reference \citenum{Tang2014}. (B) HAADF images of vortex domains in BiFeO$_3$. Reproduced with permission from Reference \citenum{Nelson2011}.}
\label{bto_pto2}
\end{figure}

Some of the most widely studied thin film ferroelectrics are those based on the $AB$O$_3$ perovskite structure, such as BTO and PbTiO$_3$ (PTO). In these systems Ti $3d$ -- O $2p$ hybridization can stabilize ferroelectric distortions,\cite{Cohen1992} leading to complex phase transitions and ferroelectric domain structures. Early research in this area employed scanning electron microscopy and TEM diffraction to visualize domain structures in BTO.\cite{HU1986} Domain structures were also explored in other systems such as PZT,\cite{Tsai1992} but resolution was still a limiting factor. With the advent of aberration-correction, stacking fault defects could be imaged in BTO using a fixed-beam TEM setup,\cite{Jia2004, Jia2003} providing direct feedback to synthesis efforts. However, spherical aberration-correction in the STEM unlocked the ability to resolve direct atomic positions in the HAADF mode and measure associated EELS spectra for interfaces such as BTO / STO\cite{KURATA2009} and BTO / Fe,\cite{Bocher2012}, as shown in Figure \ref{bto_pto}. With the ability to resolve atomic positions, it became possible to examine domain wall structures in these systems,\cite{Tang2014,Tsuda2013} which form to minimize their overall free energy, as shown in Figure \ref{bto_pto2}.A for PTO. Rhombohedral domain structures in BTO have been explored using STEM diffraction\cite{Tsuda2013} and imaging complemented with phase field simulations for PTO.\cite{Li2017}

Perhaps no multiferroic has attracted as much recent attention as BiFeO$_3$ (BFO), whose simultaneous room-temperature antiferromagnetic and ferroelectric character were shown to be enhanced through thin film deposition in the early 2000s.\cite{Wang2003} At room-temperature BFO possesses a rhombohedral $R3c$ point group ($a_{rh} = 3.965$ \AA / $\alpha_{rh} = 89.3-89.4^{\circ}$) and a perovskite-type unit cell, with Bi$^{3+}$ ions at eight-fold coordinated sites and Fe$^{3+}$ ions at six-fold coordinated sites.\cite{Palewicz2010, Moreau1971, Kubel1990} The size mismatch between the cations and oxygen leads to octahedral buckling\cite{Shannon1976a, Goldschmidt1926} and tilt of $\omega = 11 - 14^{\circ}$ around the [111] polarization direction.\cite{Palewicz2007, Moreau1971, Kubel1990} In this system, ferroelectric domain wall structures can form in 71$^{\circ}$, 109$^{\circ}$, and 180$^{\circ}$ configurations to minimize bound surface charge.\cite{Wang2013} HAADF analysis was able to directly measure off-centering of cation species relative to the oxygen sublattice, revealing the presence of complex domain structures analogous to magnetic domain walls, as shown in Figure \ref{bto_pto2}.B.\cite{Nelson2011} The directly interpretable nature of the incoherent HAADF image can be exploited to determine cation positions with picometer-level sensitivity because of the added spatial resolution afforded by spherical aberration-correction. From this, it is possible to calculate a displacement vector \textbf{D}$_{\textrm{FB}}$ for each unit cell and a local polarization \textbf{P}$_{yz}$ using the equation,

\begin{equation}
	\textbf{P}_\textrm{yz} = -2.5 \frac{\mu C}{cm^2\cdot pm}\cdot \textbf{D}_\textrm{FB}
\end{equation}

\noindent where the coefficient is estimated from Born effective charges.\cite{Neaton2005} In this way, the authors calculated the local polarization in the projected plane of the microscope image. These measurements were used as inputs for \textit{ab initio} calculations, which revealed a progression in domain wall energy of $\gamma_{109} < \gamma_{180} < \gamma_{71}$ that helps explain the resulting distribution of domain types.\cite{Wang2013} Krishnan \textit{et al.}\cite{Krishnan2015} also examined electronic structure changes associated with transitions from tetragonal to rhombohedral phases in BFO. They employed EELS measurements interpreted through extensive DFT calculations, which allowed them to fingerprint the contributions of lattice strain and $B$-site cation displacements to O $K$ edge spectral changes. This work is notable for its early use of neural networks to train the theoretical model based on experimental data (peak positions, areas, etc.).

\begin{figure}
\includegraphics[width=0.7\textwidth]{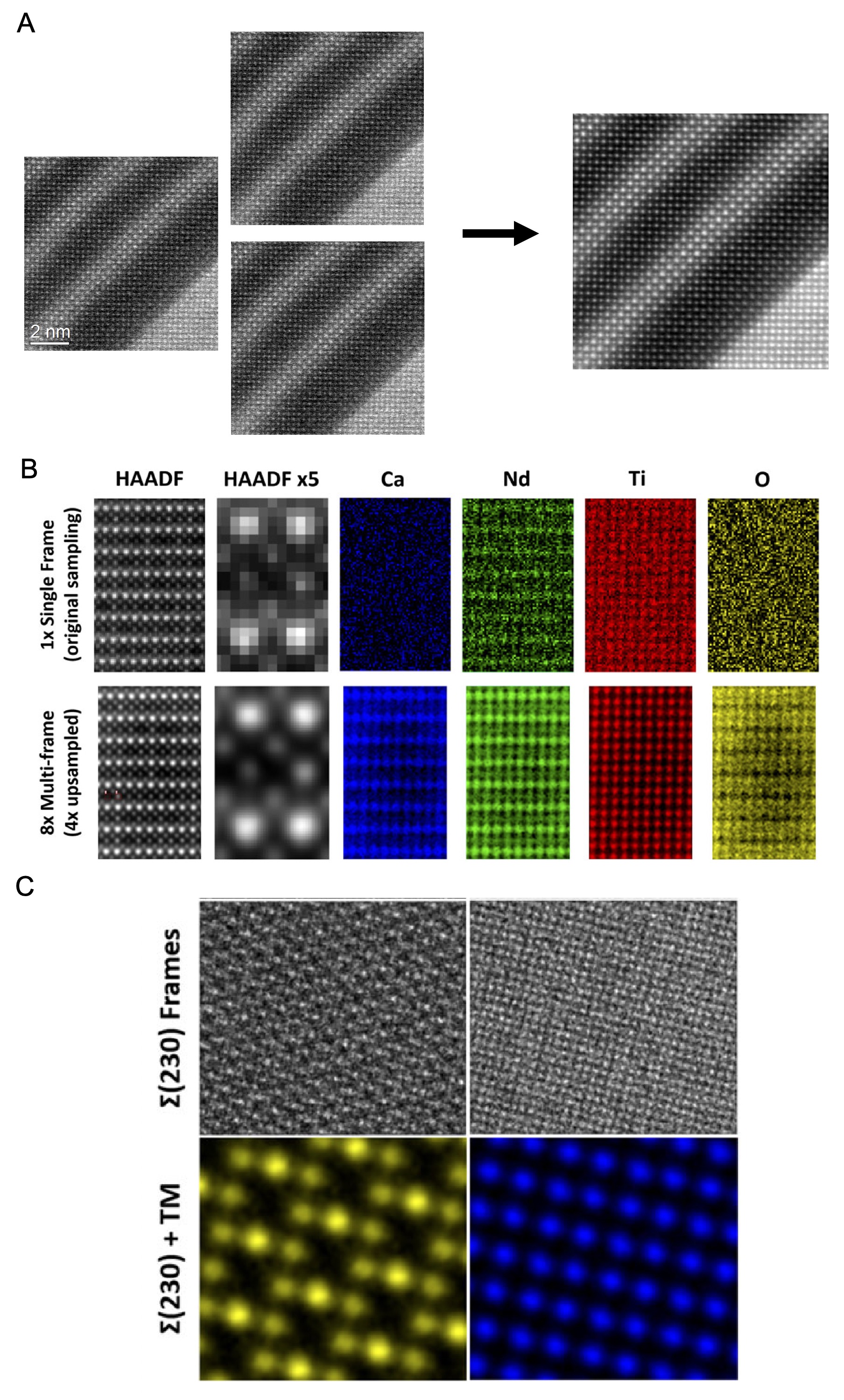}
\caption{(A) Multi-frame image acquisition and alignment process to improve HAADF signal-to-noise. Reproduced with permission from Reference \citenum{Comes2016}. (B--C) Comparison of single- vs. multi-frame spectral image acquisition showing the significant improvement in resolution and signal-to-noise. Reproduced with permission from Reference \citenum{Jones2018}.}
\label{multi-frame}
\end{figure}

Polarization analyses have also been performed on other systems, such as PZT\cite{Jia2007} and STO / LaCrO$_3$,\cite{Comes2016} with more recent studies\cite{Comes2016, Spurgeon2015} utilizing scan drift correction methods to further improve precision of the displacement measurements. Scanning probe data is acquired in a serial fashion, where the probe is rastered from point to point across a sample. This means that any deviations in sample position due to thermal drift or noise due to stray magnetic fields or vibration can limit resolution by effectively blurring out atomic column positions.\cite{Jones2013} Since the typical time for acquiring a high-resolution frame is on the order of minutes, there is plenty of opportunity for these artifacts to occur. Recently, codes have been developed to acquire many fast frames and subsequently align them using rigid and non-rigid registration routines, as shown in Figure \ref{multi-frame}.A.\cite{Jones2015,Sang2014} This method of data acquisition improves the quality of resulting images by both removing drift and reducing Poisson noise, which scales as $\sqrt{n}$, where $n$ is the number of frames averaged. This approach has the other added benefit of distributing the electron dose over many frames, allowing for beam-sensitive materials to recover from possible damage between frames. It has been extended to spectroscopic imaging\cite{Jones2018} and has enabled high-resolution chemical mapping of systems such as Nd$_{0.6}$Ca$_{0.1}$[ ]$_{0.3}$TiO$_3$ ([ ] denotes cation deficiency), shown in Figure \ref{multi-frame}.B. Building on this method, a process called template matching subdivides a single large image or spectral data set into smaller motifs, that are detected, aligned, and averaged.\cite{Jones2018} For homogeneous materials or repeating interfaces this approach can further improve signal-to-noise, particularly for low contrast atomic species (such as oxygen) or low impurity concentrations. The results can be quite dramatic for techniques like EDS, which have an inherently poor detection efficiency (just a few \% of emitted X-rays), as shown in Figure \ref{multi-frame}.C. Collectively, these methods show considerable promise for the analysis of domain wall structures, offering both improved precision and less risk of imparting beam damage effects that could change the domain wall morphology.

Microscopic domain wall analyses have also been integrated with other techniques, such as piezoresponse force microscopy (PFM), which is a largely surface sensitive method to resolve ferroelectric domain structures and functionality.\cite{Kholkin2007} High-resolution information about interface structure and film quality in BFO / La$_x$Bi$_{1-x}$FeO$_3$ multilayers,\cite{Chen2017a} for example, has helped to interpret the distribution of domain wall types measured by PFM. Screening effects in these superlattices can change the depolarizing field in BFO, leading to an increase in $109^{\circ}$ and reduction in $71^{\circ}$ type striped domain walls measured over larger areas via PFM. PFM also has been used to examine the conductivity of different domain wall types in BFO,\cite{Seidel2009a} where it was found that $109^{\circ}$ and $180^{\circ}$ domain walls were conductive, while $71^{\circ}$ were not. To interpret this behavior, Seidel \textit{et al.}\cite{Seidel2009a} examined lattice displacements across the domain wall boundary using HAADF, measuring distortions in Fe ion positions corresponding to a polarization perpendicular to the boundary. These measurements served as inputs for \textit{ab initio} calculations, which showed that a potential step should be present at the domain wall for the conductive types, leading to charge carrier accumulation. This combination of techniques helped reveal the local electronic character of domains in BFO and other ferroelectrics.

It was recognized early that the Fe--O--Fe bond angle in BFO has important consequences for magnetic and electronic order, as it dictates both magnetic exchange and orbital overlap.\cite{Catalan2009} Locally BFO is a $G$-type antiferromagnet, with each Fe$^{3+}$ spin surrounded by six antiparallel spins on its nearest Fe neighbors.\cite{Catalan2009} Magnetoelectric (ME) coupling to the FE polarization gives rise to a weak canted moment and a slight disordering of the $G$-type spin structure, which can be switched through manipulation of the polarization direction.\cite{Lebeugle2008, Lee2008} Several studies have examined the magnetic properties of BFO domain walls using a correlative imaging and magnetic characterization approach. Yang \textit{et al.}\cite{Yang2014} examined $90^{\circ}$ domain walls using STEM, PFM, and scanning tunneling microscopy (STM) under various biasing conditions. They then examined exchange coupling in a permalloy (Ni$_{81}$Fe$_{19}$) layer deposited atop the BFO, which indicated large changes in magnetoresistance as a function of domain wall morphology and polarization direction. Anisotropic magnetoresistance (AMR) measurements\cite{Lee2014a} have also been conducted on capacitively gated BFO samples and compared to TEM measurements, revealing the ferromagnetic character of domain walls in this material. As will be described in Section \ref{sec:magnetoelectrics}, this coupling between FE polarization and magnetism has received considerable attention for use in layered magnetoelectric materials.

\begin{figure}
\includegraphics[width=0.7\textwidth]{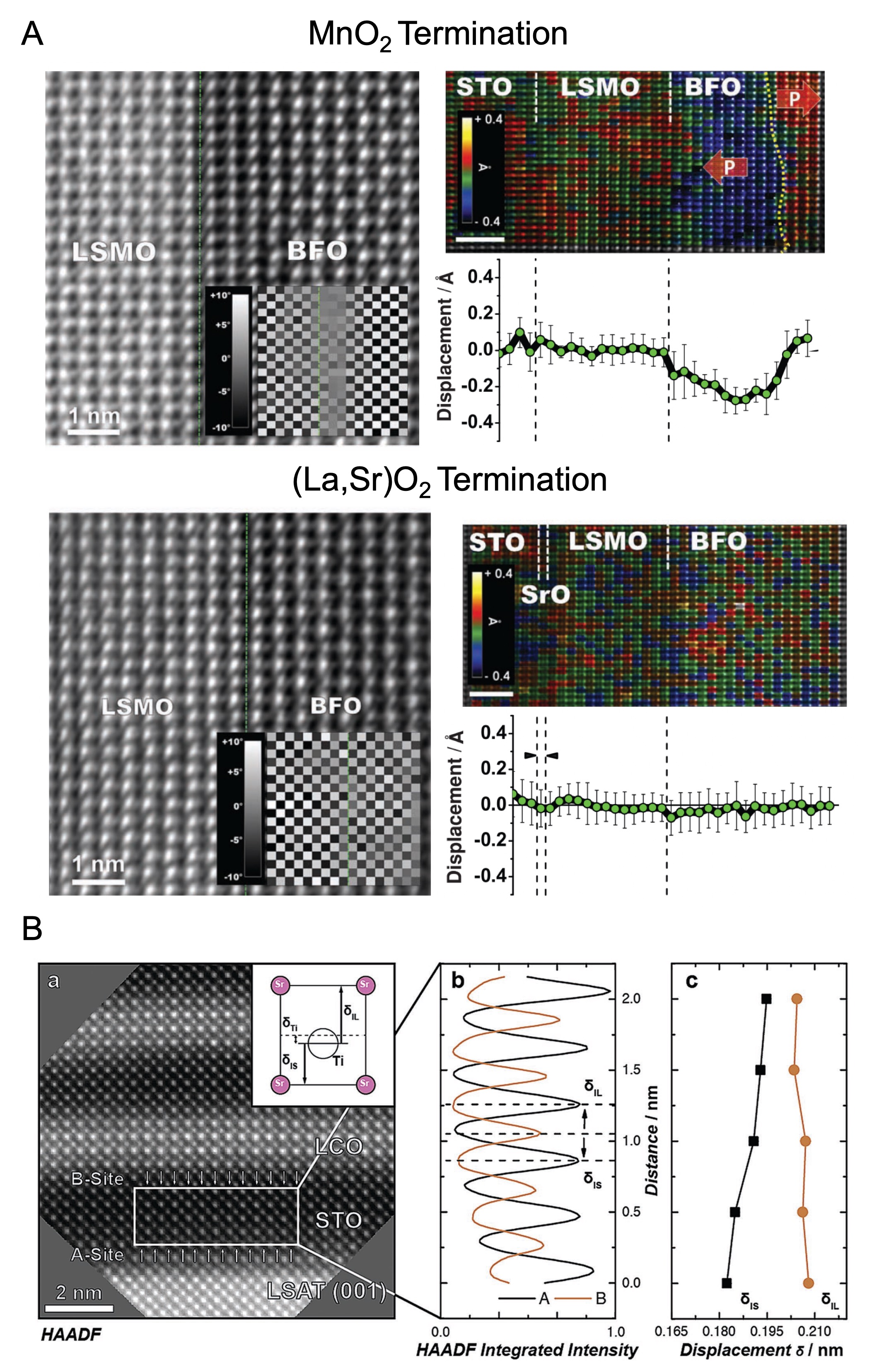}
\caption{(A) HAADF mapping of octahedral distortions and associated interfacial ferroelectric domain structures in two terminations of LSMO / BFO. Reproduced with permission from Reference \citenum{Kim2013a}. (B) Measurement of local polarization from drift-corrected HAADF imaging of asymmetric LCO / STO superlattices. Reproduced with permission from Reference \citenum{Comes2016}.}
\label{multiferroics}
\end{figure}

Finally, much work has been done to understand the role of interfaces in thin film ferroelectrics and multiferroics. As already mentioned, octahedral bond angles and tilts can significantly impact electronic and magnetic properties due to exchange coupling. Using HAADF imaging, it is possible to directly resolve changes in these bond angles at interfaces\cite{Kim2013a,Borisevich2010b} and across domain wall boundaries.\cite{Borisevich2010} Kim \textit{et al.}\cite{Kim2013a} have shown that interface termination in La$_{0.7}$Sr$_{0.3}$MnO$_3$ / BFO heterojunctions can set octahedral tilt patterns in the FE layer. In particular, for MnO$_2$-terminated LSMO, there is a complete suppression of octahedral tilts over the first 3 atomic layers, followed by gradual relaxation to bulk tilts over 3 nm. In contrast, for (La,SrO)-terminated LSMO incomplete tilt suppression occurs, which only recovers very deep ($\sim20$ nm) into the bulk of the BFO. Using peak fitting routines the authors quantified octahedral distortions at the unit cell level, as shown in Figure \ref{multiferroics}.A. They observed an oppositely-oriented, albeit ferroelectric, domain structure at the interface in the former termination, but nearly a complete anti-ferroelectric phase with zero net polarization for the latter termination. In another example, Comes \textit{et al.}\cite{Comes2016,Comes2017} constructed superlattices of polar LaCrO$_3$ and non-polar STO, engineered using MBE to create alternating positively-charged [TiO$_2$]$^0$ -- [LaO]$^+$ and negatively-charged [CrO$_2$]$^-$ -- [SrO]$^0$ layers. The resulting structure was examined at atomic-resolution using both HAADF and EELS, which indicated a remarkably sharp, well-preserved asymmetric structure across the superlattice, as shown in Figure \ref{multiferroics}.B. Moreover, drift-corrected measurements of $B$-site cation displacements indicated the emergence of a polar distortion, in agreement with predictions from theory calculations. These studies are examples of a wide body of literature that shows how interfaces can be used to tune or even completely suppress FE ordering.

\subsection{Magnetoelectric Heterostructures}\label{sec:magnetoelectrics}

While closely related to multiferroics, the special class of magnetoelectrics has a rich history in the context of thin film epitaxy and STEM characterization, so we consider it separately here. Multiferroics are generally defined as materials that simultaneously possess two or more kinds of ferroic order---ferroelectricity, ferromagnetism, or ferroelasticity---while magnetoelectrics are materials that exhibit coupling between magnetic and electrical order parameters, regardless of their nature.\cite{Eerenstein2006} These materials are of interest for both practical and fundamental reasons. Many kinds of logic devices have been proposed, including magnetic tunnel junctions (MTJs) with a magnetoelectric active layer, as well as devices in which an electric field is used to directly tune magnetization.\cite{Bea2008, Wang2010} While many single-phase magnetoelectrics have been found, including Cr$_2$O$_3$,\cite{Dzyaloshinskii1959, Astrov1960, Rado1961} GaFeO$_3$,\cite{Rado1964} and PbFe$_{0.5}$Nb$_{0.5}$O$_3$,\cite{Watanabe1989} they all suffer from a rather weak coupling between polarization and magnetization.\cite{Fiebig2005} The lack of room-temperature, single-phase magnetoelectrics has motivated the pursuit of alternative materials systems, particularly those consisting of engineered, layered structures coupled at interfaces.\cite{Srinivasan2010} In these ``artificial'' magnetoelectrics, each component system possesses disparate functionality (for example, ferromagnetism and ferroelectricity), which interact at the interface to generate unique composite behavior. Extensive work has been done to both identify and harness interfacial coupling mechanisms.

Magnetism in transition metal oxides is mediated by exchange interactions through the hybridization of O $2p$ orbitals with metal $3d$ cations, giving rise to predominantly antiferromagnetic (super-exchange) or ferri/ferromagnetic (double-exchange) order. Magnetic order and local polarization in these systems are closely related, \cite{Lee2008, Zhao2006} as described in Section \ref{sec:ferroelectrics}. In particular, extensive work has focused on interfacial control of magnetization through the exchange bias interaction, using materials such as multiferroic BFO. When a bilayer composite of antiferromagnetic and ferromagnetic materials is heated above the N\'eel temperature of the AF and then cooled in an external magnetic field, AF spins at the interface act to pin adjacent FM spins.\cite{Kiwi2001} This interaction imposes an extra coercive force on the FM layer, leading to a shift in hysteresis behavior ($H_{EB}$). It has been shown that magnetization may be tuned through the exchange bias effect by coupling BFO to an adjacent FM; then, by switching the FE polarization, one can induce changes in this bias.\cite{Wu2010}

\begin{figure}
\includegraphics[width=\textwidth]{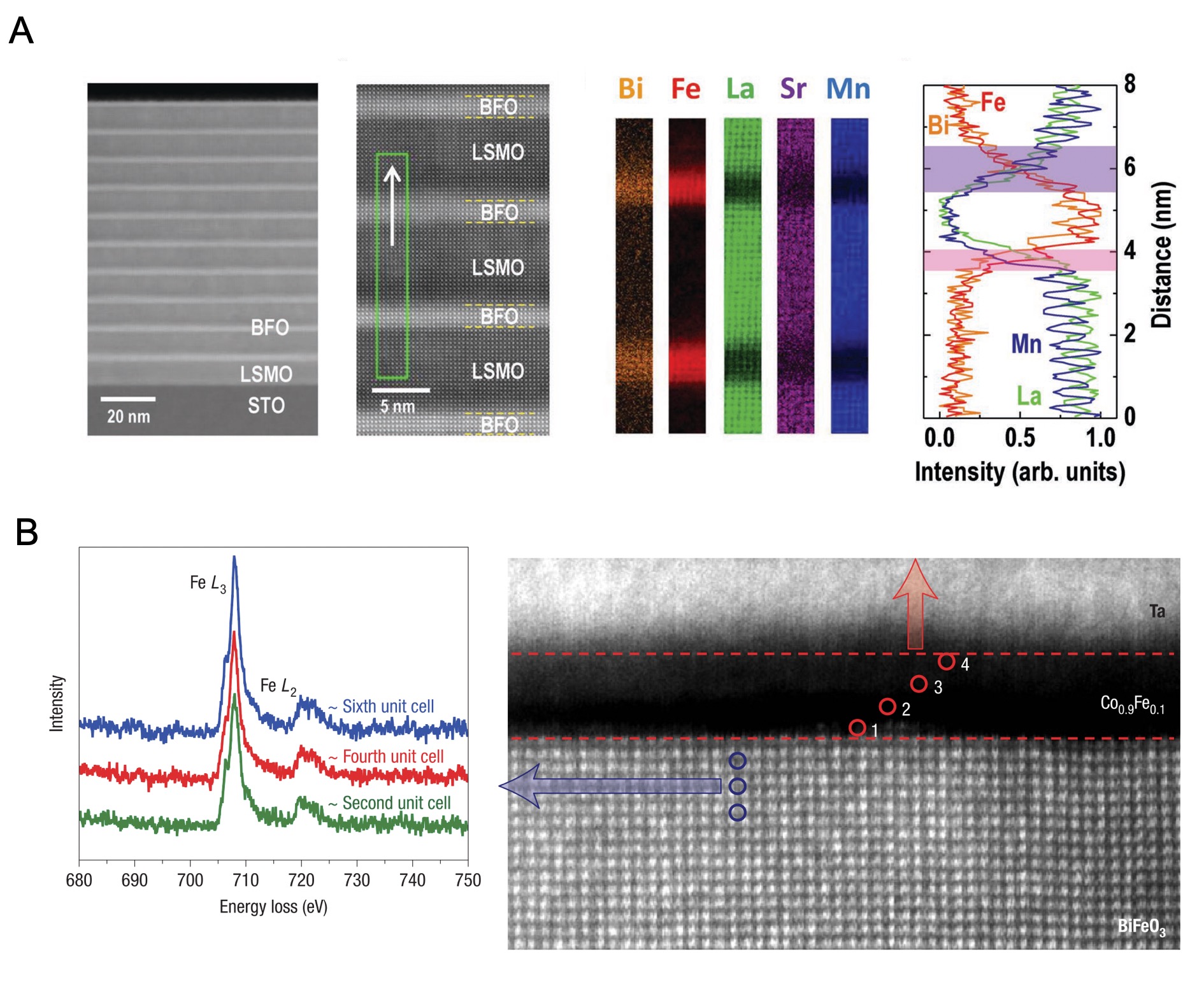}
\caption{(A) HAADF and EELS examination of BFO / LSMO structures exhibiting exchange-bias-mediated coupling. Reproduced with permission from Reference \citenum{Guo2017}. (B) EELS analysis of CoFe / BFO interfaces showing a lack of oxidation or structural defects. Reproduced with permission from Reference \citenum{Chu2008}.}
\label{eb-coupling}
\end{figure}

BFO coupling to manganites such as La$_{0.7}$Sr$_{0.3}$MnO$_3$\cite{Wu2013a,Wu2010} has received considerable attention because of their good structural compatibility and potential for a well-matched, high-quality interface. Guo \textit{et al.}\cite{Guo2017} have examined LSMO / BFO superlattices using a combination of analytical STEM and polarized neutron reflectometry (PNR). As shown in Figure \ref{eb-coupling}.A, they examined the structure and chemistry of the film layers, revealing minimal intermixing less than 1--2 unit cells. This information was used to refine the PNR data to extract magnetic depth profiles, which showed that orbital ordering between Fe and Mn sites leads to a strong canted magnetization in the BFO. Importantly, through local imaging the authors were able to rule out other effects, such as charge transfer, intermixing, epitaxial strain, and orbital rotations/tilts. Coupling between BFO and ultrathin ferromagnetic Co$_{0.9}$Fe$_{0.1}$ (CoFe) has also been studied by Chu \textit{et al.} \cite{Chu2008} using a multimodal approach in which BFO domain structures were switched using PFM and changes in magnetic domain structures were measured using XMCD--photoemission electron microscopy (PEEM). The authors found that the magnetic domains rotated 90$^{\circ}$ upon application of an electric field to BFO, which could be reversed by removal of the field. In this work EELS fine structure measurements confirmed the absence of defects or oxidation at the CoFe / BFO interface that would interfere with coupling, as shown in Figure \ref{eb-coupling}.B, helping to validate coupling between uncompensated interfacial spins in the structure.

Another well-studied mechanism, charge-mediated coupling in oxide heterostructures finds its origin in semiconductor field-effect devices, such as metal-oxide-semiconductor field-effect transistors (MOSFETs).\cite{Ahn2006} The areal carrier density for these device lies in the range of $10^{12}-10^{13}$ charges cm$^{-2}$ for a 10 nm channel, which can be tuned by a gate dielectric such as SiO$_2$.\cite{Ahn2006} However, FEs such as PZT offer a remanent polarization of $3 \times 10^{14}$ charges cm$^{-2}$, which is an order of magnitude larger than the breakdown field of SiO$_2$.\cite{Hong2005} This property indicates that an FE layer may be used to induce sizable modulations of the carrier density in a metal. It should be noted that these modulations are screened quite quickly by free carriers in a metal. One can estimate the expected Thomas-Fermi (TF) screening length ($\lambda_{TF}$) in a metal using the equation, 

\begin{equation}
\lambda_{TF} = \sqrt{\varepsilon b / (4 \pi e^2 \partial n / \partial \mu)}
\end{equation}

\noindent where $\varepsilon$ is the background dielectric constant of the oxide, $b$ is the interplanar spacing, and $n(\mu)$ is the chemical potential dependence of the charge carrier density.\cite{Ahn2006} According to this equation, the expected screening length for most metals is on the order of a few u.c., largely confining this effect to a small layer at an interface.

In the case of charge-coupled ME composites, this field effect is generally used in one of three ways: to directly modify the magnetic moment of a system, to change the magnetic interactions present in the system, or to change the magnetic anisotropy in the system.\cite{Vaz2012} In the first case, charge-screening by the metal modifies the spin asymmetry at the Fermi level, giving rise to a change in magnetic moment.\cite{Zhang1999} This has been predicted theoretically in several heterostructures, including Fe / BTO (001) and Fe$_3$O$_4$ / BTO, where the ferroelectric causes local bonding changes at the interface.\cite{Niranjan2008, Fechner2008} It has also been observed in the manganites, particularly La$_{0.8}$Sr$_{0.2}$MnO$_3$ / Pb(Zr$_{0.8}$Ti$_{0.2}$)MnO$_3$, where the bound surface charge from the FE directly affects the adjacent Mn valence.\cite{Vaz2010a} 

\begin{figure}
\includegraphics[width=\textwidth]{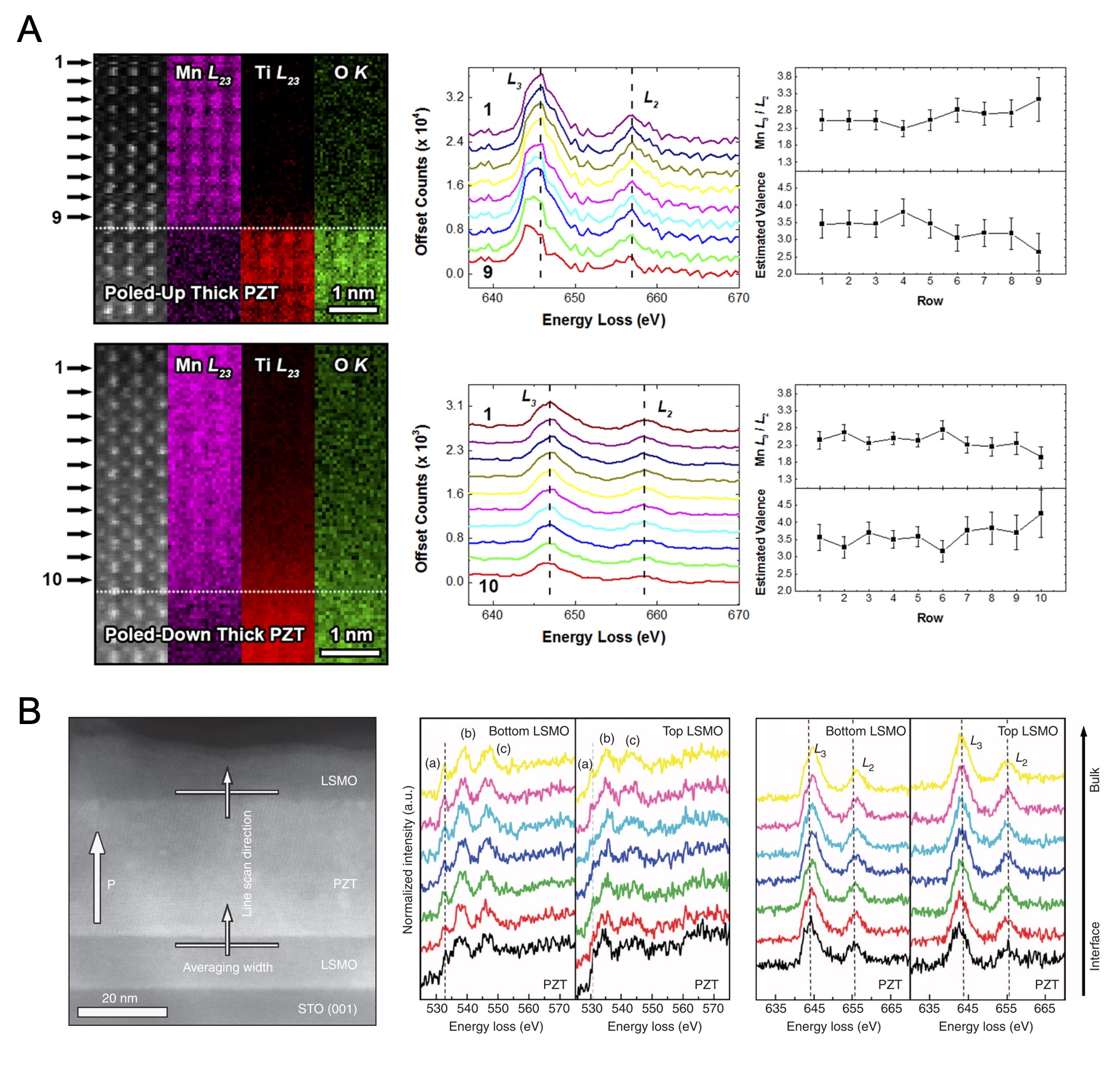}
\caption{(A) EELS measurements of polarization-dependent change in effective local Mn valence in PZT / LSMO heterostructures. Reproduced with permission from Reference \citenum{Spurgeon2014}. (B) EELS measurements of changes in O $K$ and Mn $L_{2,3}$ fine structure resulting from asymmetric poling in PZT / LSMO. Reproduced with permission from Reference \citenum{Spurgeon2015}.}
\label{charge-coupling}
\end{figure}

The second mechanism of charge-mediated coupling tunes the magnetic interactions present in a system. Bound surface charge from a FE can modulate the carrier density in these compounds and stabilize FM order. Even in the case of metallic complex oxides, such as LSMO, it is possible to induce sizable modulation of carrier densities, albeit across a shorter screening length.\cite{Vaz2010a, Vaz2010b} In these compounds doping of La by a divalent alkaline earth, such as Sr or Ca, removes an $e_g$ carrier from the system and leads to a transition from Mn$^{3+}$ to Mn$^{4+}$ valence---analogous to hole doping.\cite{Vaz2012} Spurgeon \textit{et al.}\cite{Spurgeon2014} studied screening affects at the LSMO / PZT interface, using atomic-scale EELS to examine the local Mn charge state. As shown in Figure \ref{charge-coupling}.A, they observed a deviation from the bulk Mn valence of $\sim3.4$ that appeared to depend on PZT polarization direction and operated over $<2$ nm, consistent with the expected TF screening length. These results were compared to depth-resolved magnetization measurements that revealed sizable differences for the two states. In addition, in LSMO / PZT the interfacial hole charge depletion state gives rise to FM ordering and an accumulation state gives rise to AF ordering at low-temperature.\cite{Vaz2010a, Vaz2011} Vaz \textit{et al.} proposed that such a spin structure change is necessary to account for the induced magnetization change upon switching the PZT polarization direction; they cited several first principles studies of related LSMO / STO and La$_{0.5}$Ba$_{0.5}$MnO$_3$ / BTO systems, in which it is calculated that an AF-$A$-type configuration represents the lowest energy ground state.\cite{Burton2009, Fang2000} Spurgeon \textit{et al.}\cite{Spurgeon2015} also explored this behavior using local chemical mapping and \textit{ab initio} calculations in symmetric LSMO / PZT / LSMO interfaces. They were able to map the local change in effective Mn valence and expected magnetic phase, finding a suppression of local ferromagnetic ordering, albeit of differing magnitude for the two interfaces. Their results showed that local changes in EELS fine structure can be correlated back to both the Mn valence and spin state, providing a route to map magnetization in the STEM, as shown in Figure \ref{charge-coupling}.B. More recent work\cite{Soni2016,Meng2019} has observed similar changes in this system.

The third mechanism of charge-mediated coupling occurs when charge screening modifies the magnetocrystalline anisotropy (MCA) of a system. This mechanism is intimately connected to the previous two mechanisms, since a reduction in magnetic moment will affect the magnetostatic energy and a change in the exchange interactions of the system will affect the domain wall formation energy.\cite{Vaz2012} Moreover, it is expected that a change in orbital occupancy will also change the MCA of a system. These kinds of changes have been demonstrated in various systems, including FePt and FeCo, as well as in ultrathin Fe films;\cite{Weisheit2007, Maruyama2009} very thin layers are more likely to exhibit such coupling, since surface MCA will dominate their behavior.\cite{Coey2009} Parkin \textit{et al.}\cite{Parkin2004a} examined CoFe / MgO tunnel barriers in which anisotropy can be tuned by an applied electric field, using high-resolution TEM imaging to confirm the quality and epitaxy of the layers in the structure to support magnetic measurements. In summary, while there are various mechanisms of charge-mediated coupling, they are all generally constrained to thin interface layers, whose quality and epitaxy are well assessed using highly local structural probes.

\subsection{Synthesis Pathways in Complex Oxides}

The examination of synthesis pathways in complex oxides represents one of the most broad and important use cases for STEM techniques. Because of the highly non-equilibrium nature of thin film growth, desired and realized structures often diverge. Substrate effects, variations in thin film growth rates and oxygen pressure, and local chemical fluctuations can all lead to unintentional structural and chemical deviations. Understanding and controlling these defects depends (1) on an awareness of their existence, (2) appropriate theory models for growth processes and defect formation, and (3) a way to link experimental observations to theory. While volume- and area-averaged X-ray and neutron scattering methods are powerful techniques to examine the global properties of materials, they can be less sensitive to infrequent or aperiodic defects. Direct structural and chemical imaging can fill in these gaps, providing complementary local insights into the growth behavior of oxides.

\begin{figure}
\includegraphics[width=\textwidth]{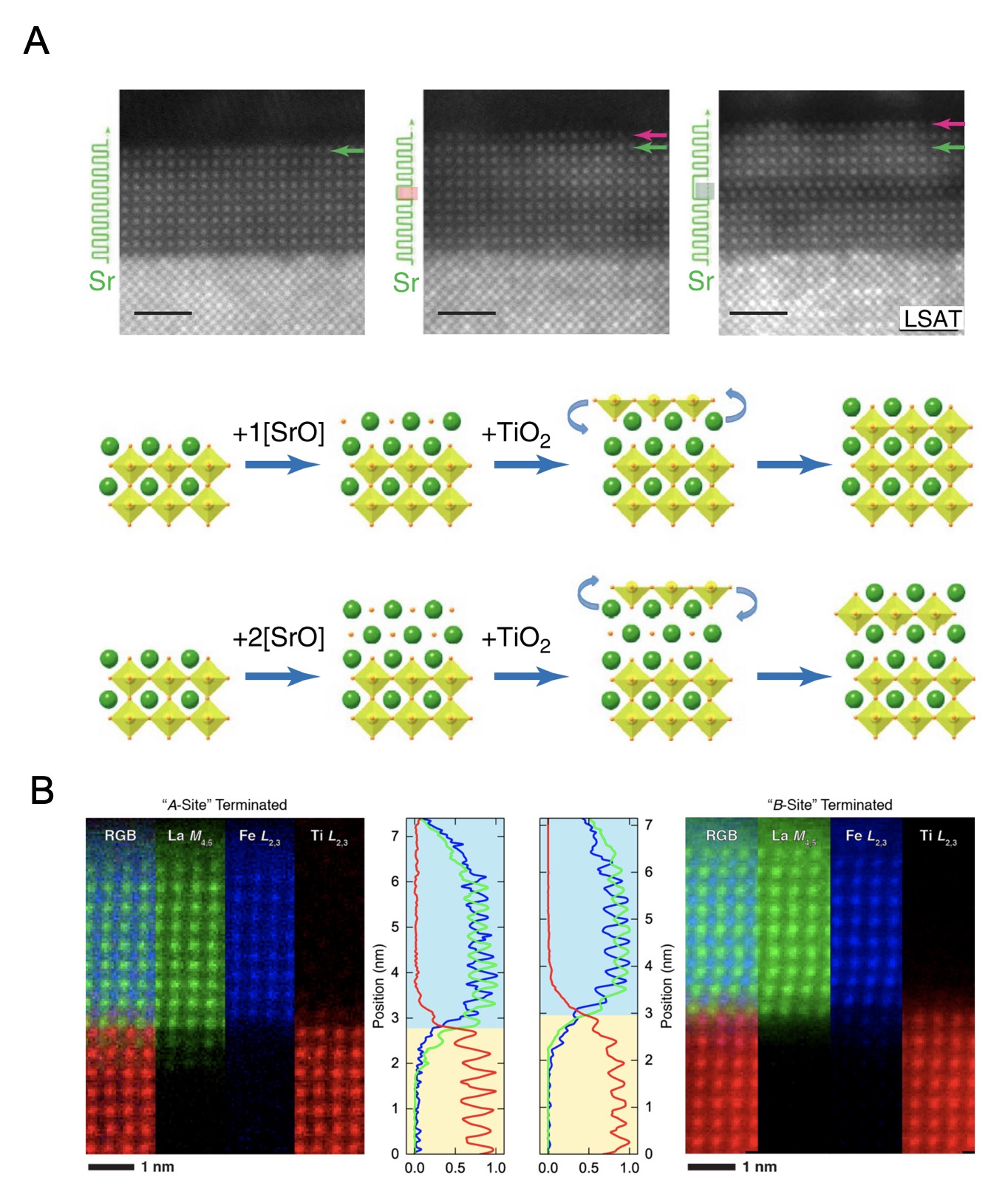}
\caption{(A) HAADF analysis of floating layers in STO / LSAT. Reproduced with permission from Reference \citenum{Nie2014a}.  (B) EELS chemical mapping of interfacial rearrangement in LFO / STO heterojunctions. Reproduced with permission from Reference \citenum{Spurgeon2017}. }
\label{rearrangements}
\end{figure}

Understanding the reorganization of crystal layers during MBE growth is essential to control  interface configurations and achieve desired functionality. Work by Lee \textit{et al.}\cite{Lee2014} and others has shown that layered oxide materials can dynamically rearrange during high-temperature growth of Sr$_2$TiO$_4$ / STO interfaces. Using \textit{in situ} synchrotron diffraction and theory calculations, they determined that a significant thermodynamic driving force for layer rearrangement can occur during growth, transforming a [SrO] -- [SrO] -- [TiO$_2$] configuration to [SrO] -- [TiO$_2$] -- [SrO]. Similar observations have been made by Nie \textit{et al.},\cite{Nie2014a} who found that layer-by-layer growth film growth can be subject to complex monolayer rearrangements that must be accounted for to achieve desired targets. As shown in Figure \ref{rearrangements}.A, using HAADF imaging they were able to discern the presence of a [SrO] layer that had migrated to the surface of a STO film grown on LSAT, in spite of their attempts to engineer an [SrO] double layer into the film. They found that this layer was robust and appeared to ``float'' to the top of the sample during growth. Interface rearrangements were also observed by Spurgeon \textit{et al.}\cite{Spurgeon2017} in layer-by-layer growth of LaFeO$_3$ / STO interfaces. The authors synthesized nominally [SrO]- and [TiO$_2$]-terminated STO homoepitaxial films, as confirmed by XPS. However, after deposition of LFO, the resulting heterojunctions appeared to rearrange to identical configurations, as confirmed by the EELS results shown in Figure \ref{rearrangements}.B. They compared their experimental results to theory calculations, which found that the [SrO] -- [FeO$_2$] interface configuration was energetically unfavorable, likely leading to its disappearance. These results helped rationalize prior results that showed very similar band offsets and potential gradients for the two interfaces.\cite{Comes2016a} These and other studies\cite{Andersen2018} underscore the important roles of thermodynamic and kinetic limitations in achieving desired interface configurations.

\begin{figure}
\includegraphics[width=0.9\textwidth]{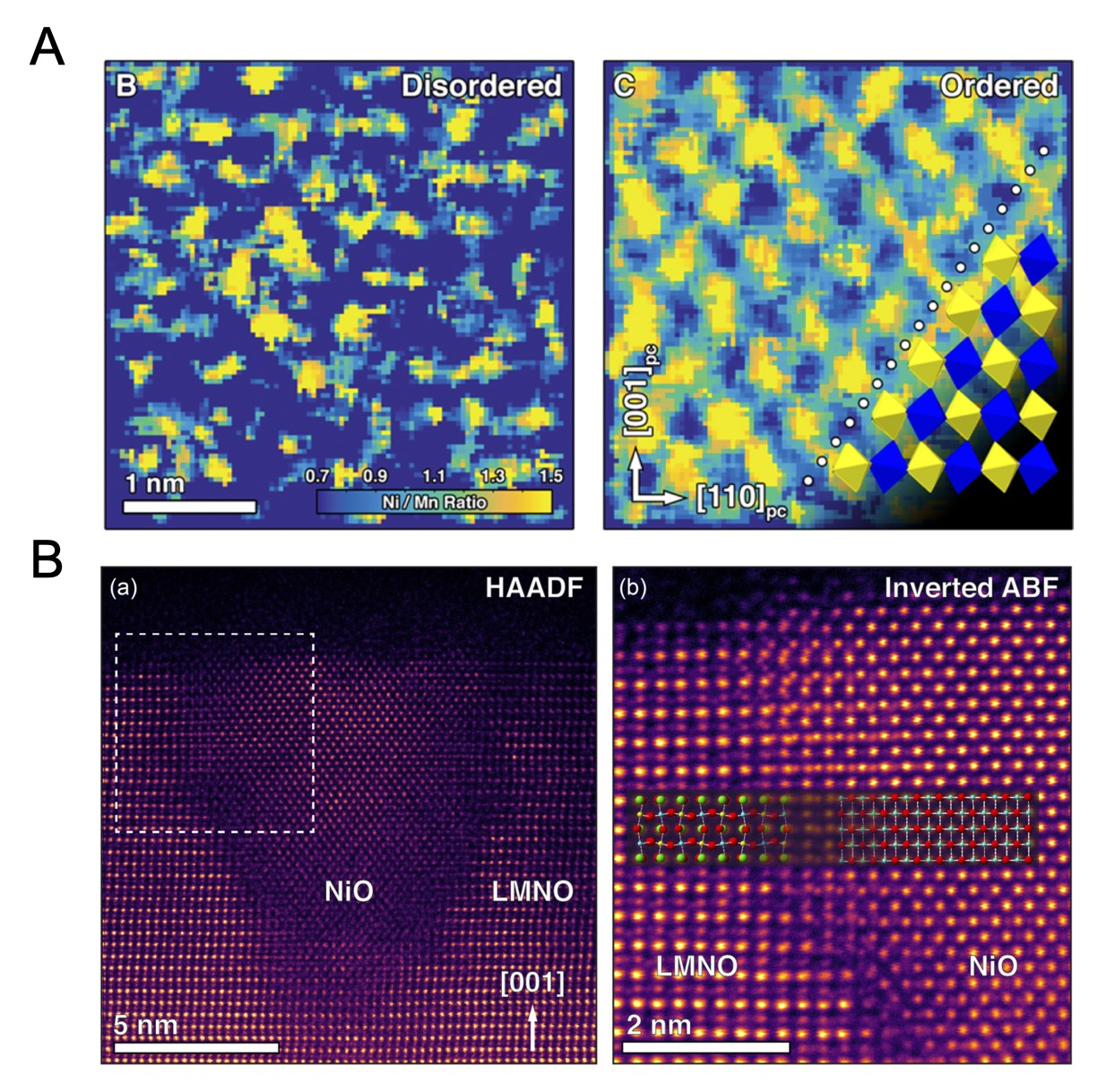}
\caption{(A) Atomic-scale STEM-EDS analysis of annealing-induced cation ordering in LMNO. Reproduced with permission from Reference \citenum{Spurgeon2016}. (B) HAADF and ABF imaging of nanoscale NiO inclusions in LMNO. Reproduced with permission from Reference \citenum{Spurgeon2018}.}
\label{phase-sep}
\end{figure}

\begin{figure}
\includegraphics[width=0.8\textwidth]{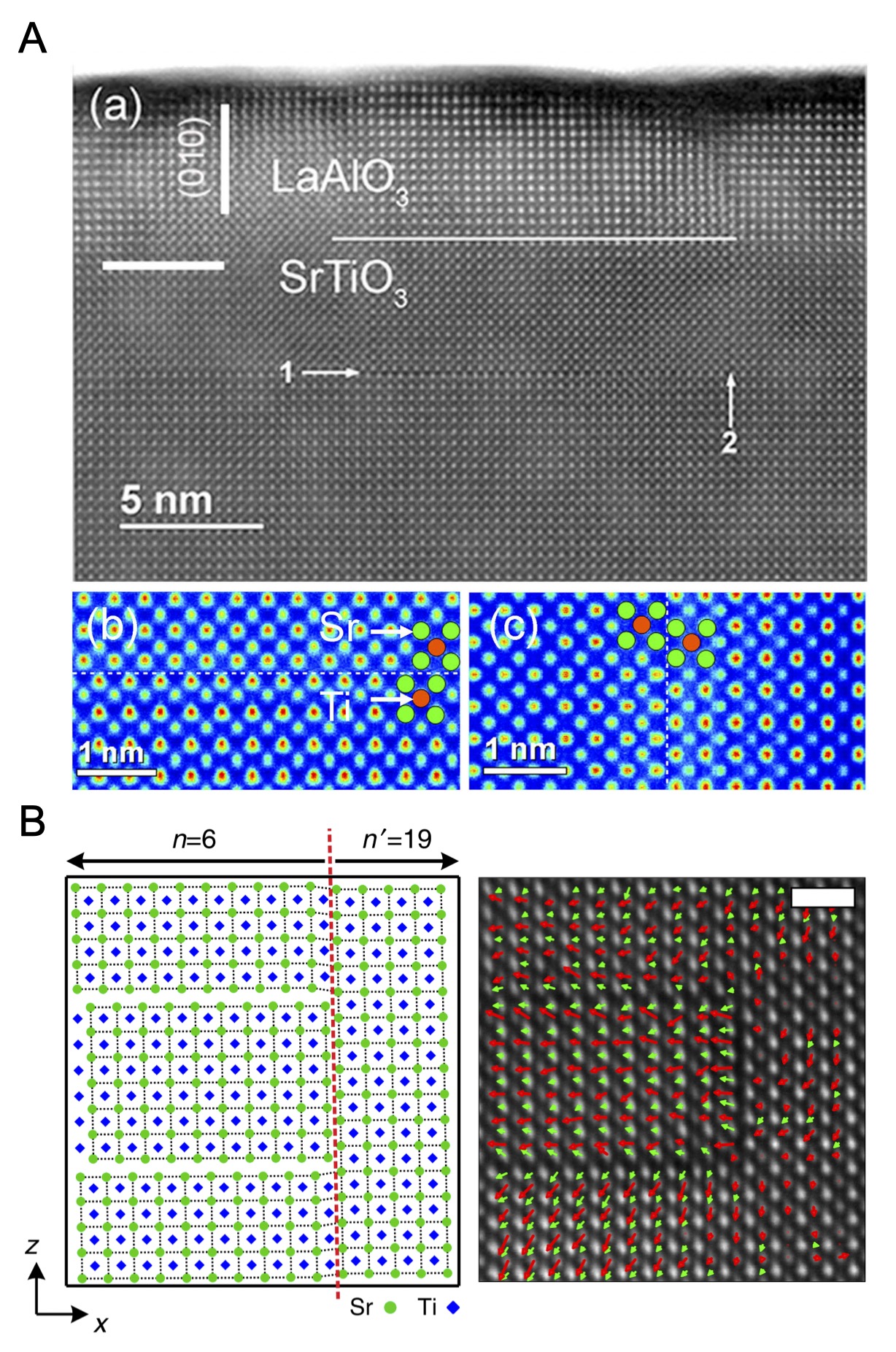}
\caption{(A) HAADF imaging of formation of APBs in Sr-rich STO. Reproduced with permission from Reference \citenum{Xu2016}. (B) Polar distortions measured from HAADF images in RP phases due to local structural rumpling. Reproduced with permission from Reference \citenum{Stone2016}.}
\label{phase-sep2}
\end{figure}

Beyond rearrangements at interfaces, the resulting phase distribution in a film structure can be disrupted by minor changes in growth parameters, as well as kinetic and thermodynamic limitations. Measurement of these phases, which can be quite non-uniform and on the length scale of just a few nanometers, is challenging using most lab-based X-ray sources. These dimensions are well below the lateral coherence length of X-rays ($\sim10-100$ \textmu m), necessitating the use of more local probes. Such nanoscale phase separation has been examined in double perovskite materials, in which achieving cation ordering for favorable magnetic properties is challenging using low $p_{O_2}$ MBE deposition. Spurgeon \textit{et al.}\cite{Spurgeon2018,Spurgeon2016} have examined MBE-grown La$_2$MnNiO$_6$ (LMNO), whose nominal stoichiometry and crystal structure was optimized, as confirmed by volume-averaged XRD techniques. In spite of this, the authors found that the measured magnetic moment of the as-grown film was substantially less than the expected value of 5.0 \textmu$_\textrm{B}$ f.u.$^{-1}$. Upon annealing this moment improved, but still did not approach the theoretical maximum. Atomic-scale STEM-EDS measurements showed an improvement in local cation ordering in the lattice upon annealing, but also indicated the presence of disordered structural regions, as shown in Figure \ref{phase-sep}.A. Subsequent analysis showed these regions to be nanoscale-sized inclusions of NiO (Figure \ref{phase-sep}.B) that had phase separated from the matrix, reducing its effective ferromagnetic moment. Theory calculations based on this data indicated that an oxygen-poor growth environment, coupled with interface charge effects, are the likely driver for this behavior. Similar kinds of structural inclusions have been observed in Ruddlesden-Popper phases by Xu \textit{et al.}\cite{Xu2016} during growth of Sr-rich STO. These subtle defects, shown in Figure \ref{phase-sep2}.A, appear to form due to kinetic limitations, resulting in anti-phase boundaries (APBs) when film overgrowth occurs around them. Other such changes have been observed in Sr$_7$Ti$_6$O$_{19}$, another RP phase.\cite{Stone2016} In this material it was found that rumpling of rocksalt blocks in the structure can lead to associated polar distortions, as measured by correlative HAADF and ABF imaging shown in Figure \ref{phase-sep2}.B. These results were compared to \textit{ab initio} calculations, which revealed that highly-strained layered and defective RP phases can host a variety of small nanoscale structural distortions. Because of their small spatial dimensions, it is difficult to access and measure these structures in any other way.

\begin{figure}
\includegraphics[width=0.9\textwidth]{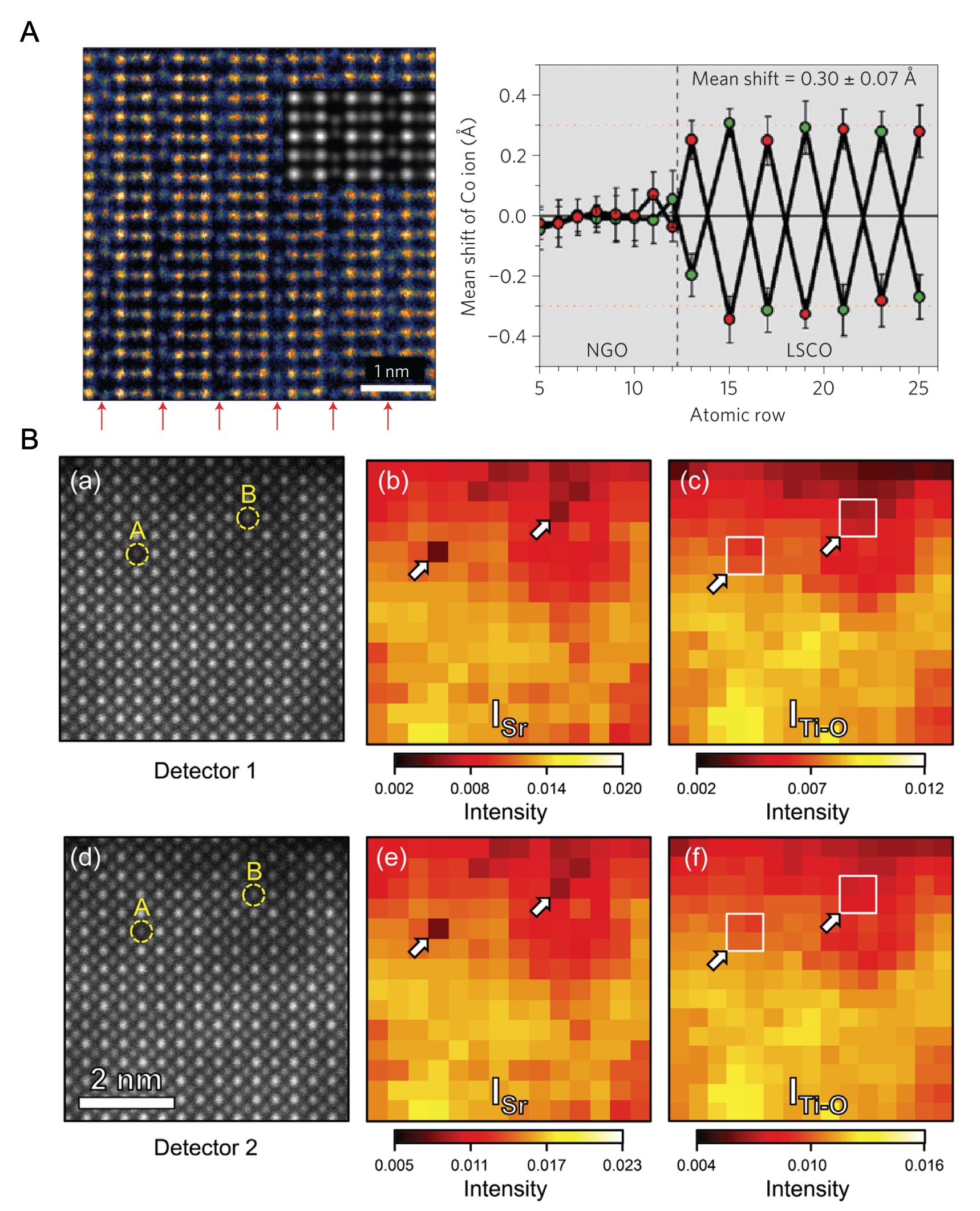}
\caption{(A) HAADF image and corresponding fits to the lattice positions, used to detect the presence of oxygen vacancies. Reproduced with permission from Reference \citenum{Kim2012b}. (B) ADF imaging of Sr vacancies in Sr-deficient STO. Reproduced with permission from Reference \citenum{Kim2016}.}
\label{point-defects}
\end{figure}

A final important consideration is the presence of point defects, such as oxygen vacancies, which have a large effect on the properties of complex oxide interfaces. These structures can form in response to local fluctuations during growth and result from film-substrate interactions. Typically these defects are examined using non-local methods, such as electrical transport or XRD (if they are sufficiently ordered). However, it is possible to examine point defects directly in the STEM. For example, Kim \textit{et al.}\cite{Kim2012b} have examined variations in lattice spacing in Brownmillerite La$_{0.5}$Sr$_{0.5}$CoO$_{2.5}$ using HAADF. As shown in Figure \ref{point-defects}.A, they observe minor shifts in periodic lattice spacing in the vicinity of the film-substrate interface. Through comparison to theory, they found that the underlying octahedral tilt patterns of the substrate can drive changes in oxygen stoichiometry. In another case, Kim \textit{et al.}\cite{Kim2016} examined the presence of Sr vacancies in MBE-grown STO. Using drift-corrected imaging, they detected weak changes in HAADF contrast, identifying statistically-significant changes in Sr column positions corresponding to vacancies resulting from slightly Sr-deficient growth conditions, as shown in Figure \ref{point-defects}.B. In this case, contrast case was quantified through the use of multislice image simulations, which helped establish the position in the atomic column where these defects might be located. In addition, lattice distortions around the vacancy sites were measured, revealing the response of the lattice to these defects.

\section{Summary}

As described in this chapter, much of our understanding of complex oxide interfaces and heterostructures can be attributed to STEM techniques. Modern instruments provide access to a rich variety of spatial and chemical information that informs materials synthesis, properties, and performance. STEM plays a unique role in materials science, chemistry, biology, and physics, since it provides information that is difficult or impossible to access using any other single characterization technique. Importantly, data from the microscope are highly complementary to other experimental and theoretical methods, benefitting from strong interpretive frameworks that allow us to extract deep predictive insight. The future will continue to see innovations in resolution and sensitivity, as well as increasing integration of data science into the instrument to streamline the process of data collection and interpretation, yielding transformative new insights into functional materials systems.

\clearpage

\section*{Acknowledgments}

I would like to thank Dr. Andrew Lang for reviewing the manuscript, Dr. Colin Ophus for useful discussions, and Dr. Bethany Matthews for her assistance preparing the UO$_2$ samples shown in this manuscript. This work was supported by the U.S. Department of Energy (DOE), Office of Basic Energy Sciences, Division of Materials Science and Engineering under Award No. 10122. The UO$_2$ samples were analyzed for a project supported by the Laboratory Directed Research and Development (LDRD) Nuclear Processing Science Initiative (NPSI) at Pacific Northwest National Laboratory (PNNL). PNNL is a multi-program national laboratory operated for the DOE by Battelle. A portion of the microscopy work was conducted in the Radiological Microscopy Suite (RMS), located in the Radiochemical Processing Laboratory (RPL) at PNNL.

\clearpage

\bibliography{References}

\end{document}